\documentclass[12pt]{article}

\usepackage[utf8]{inputenc} 
\usepackage[english]{babel} 
\usepackage[T1]{fontenc}
\usepackage{amsmath} 
\usepackage{amssymb}
\usepackage{amsthm}
\usepackage[ruled]{algorithm2e}
\usepackage{bm} 
\usepackage{braket}
\usepackage{csquotes}
\usepackage{eucal} 
\usepackage{hyperref} 
\usepackage{enumitem} 
\usepackage{parskip}
\usepackage{microtype}
\usepackage{cite}
\usepackage{tikz-cd}
\usepackage[left=1in,top=1in,right=1in,bottom=1in]{geometry}

\newcommand{\BB}[1]{\mathbb{#1}}
\newcommand{\B}[1]{\bm{#1}}

\newcommand{\CC}[1]{\mathcal{#1}}
\newcommand{\SC}[1]{\textsc{#1}}

\DeclareMathOperator{\poly}{poly}
\DeclareMathOperator{\polylog}{polylog}
\DeclareMathOperator{\rank}{rank}
\newcommand{\est}{\text{est}}
\newcommand{\eps}{\varepsilon}

\newcommand{\mrainformal}{
  Suppose we are given as input a matrix $A$ supporting query and $\ell^2$-norm sampling operations, a row $i \in [m]$, a singular value threshold $\sigma$, an error parameter $\eta > 0$, and a sufficiently small $\eps > 0$.
  There is a classical algorithm whose output distribution is $\eps$-close in total variation distance to the distribution given by $\ell^2$-norm sampling from the $i$th row of a low-rank approximation $D$ of $A$ in query and time complexity
    \[ O\left(\poly\Big(\frac{\|A\|_F}{\sigma}, \frac{1}{\eps}, \frac{1}{\eta}, \frac{\|A_i\|}{\|D_i\|}\Big)\right), \]
  where the quality of $D$ depends on $\eta$ and $\eps$.
}

\newcommand{\mra}{
  There is a classical algorithm that, given a matrix $A$ with query and sampling assumptions as described in Proposition~\ref{prop:ds-matrix}, along with a row $i \in [m]$, threshold $\sigma$, $\eta \in (0,1]$, and sufficiently small $\eps > 0$, has an output distribution $\eps$-close in total variation distance to $\CC{D}_{D_i}$ where $D \in \BB{R}^{m\times n}$ satisfies $\|D - A_{\sigma,\eta}\|_F \leq \eps\|A\|_F$ for some $A_{\sigma,\eta}$, in query and time complexity
    \[ O\left(\poly\Big(\frac{\|A\|_F}{\sigma}, \frac{1}{\eps}, \frac{1}{\eta}, \frac{\|A_i\|}{\|D_i\|}\Big)\right). \]
}

\newcommand{\mrbinformal}{
  Applying Theorem~\ref{main-result-a} to the recommendation systems model with the quantum state preparation data structure achieves identical bounds on recommendation quality as the quantum algorithm in \cite{kerenidis2016quantum} up to constant factors, for sufficiently small $\eps$.
}

\newcommand{\mrb}{
  Suppose we are given $\hat{T}$ in the data structure in Proposition~\ref{prop:ds-matrix}, where $\hat{T}$ is a subsample of $T$ with $p$ constant and $T$ satisfying $\|T-T_k\|_F \leq \rho\|T\|_F$ for a known $k$.
  Further suppose that the premises of Theorem~\ref{thm:lowrank-to-approx-tech} hold, the bound in the conclusion holds (which is true with probability $\geq 1-\exp(-19(\log n)^4)$), and we have $S$ a $(\gamma, \zeta)$-typical set of users with $1-\zeta$ and $\gamma$ constant.
  Then, for sufficiently small $\eps$, sufficiently small $\rho$ (at most a function of $\zeta$ and $\gamma$), and a constant fraction of users $\bar{S} \subset S$, for all $i \in \bar{S}$ we can output samples from a distribution $\CC{O}_i$ satisfying
    \[ \|\CC{O}_i - \CC{D}_{T_i}\|_{TV} \lesssim \eps + \rho \]
  with probability $1-(mn)^{-\Theta(1)}$ in $O(\poly(k,1/\eps)\polylog(mn))$ time.
}

\makeatletter
\def\thm@space@setup{%
    \thm@preskip=\parskip \thm@postskip=0pt
}
\makeatother

\newtheorem{theorem}{Theorem}[section]
\newtheorem*{theorem*}{Theorem}
\newtheorem{mr}{Theorem}
\newtheorem*{mr*}{Theorem}
\newtheorem{proposition}[theorem]{Proposition}
\newtheorem*{proposition*}{Proposition}
\newtheorem{corollary}[theorem]{Corollary}
\newtheorem*{corollary*}{Corollary}

\theoremstyle{definition}
\newtheorem*{definition}{Definition}
\newtheorem{lemma}[theorem]{Lemma}
\newtheorem*{lemma*}{Lemma}

\newtheorem*{problem*}{Problem}

\author{Ewin Tang}
\title{A quantum-inspired classical algorithm for recommendation systems}

\begin{document}
\maketitle
\begin{abstract}
We give a classical analogue to Kerenidis and Prakash's quantum recommendation system, previously believed to be one of the strongest candidates for provably exponential speedups in quantum machine learning.
Our main result is an algorithm that, given an $m \times n$ matrix in a data structure supporting certain $\ell^2$-norm sampling operations, outputs an $\ell^2$-norm sample from a rank-$k$ approximation of that matrix in time $O(\operatorname{poly}(k)\log(mn))$, only polynomially slower than the quantum algorithm.
As a consequence, Kerenidis and Prakash's algorithm does not in fact give an exponential speedup over classical algorithms.
Further, under strong input assumptions, the classical recommendation system resulting from our algorithm produces recommendations exponentially faster than previous classical systems, which run in time linear in $m$ and $n$.

The main insight of this work is the use of simple routines to manipulate $\ell^2$-norm sampling distributions, which play the role of quantum superpositions in the classical setting.
This correspondence indicates a potentially fruitful framework for formally comparing quantum machine learning algorithms to classical machine learning algorithms.
\end{abstract}
\tableofcontents
\section{Introduction}

\subsection{Quantum Machine Learning} \label{subsec:qml}
This work stems from failed efforts to prove that Kerenidis and Prakash's quantum recommendation system algorithm \cite{kerenidis2016quantum} achieves an exponential speedup over any classical algorithm.
Such a result would be interesting because Kerenidis and Prakash's algorithm is a quantum machine learning (QML) algorithm.

Though QML has been studied since 1995 \cite{Bshouty:1999:LDO:305673.305756}, it has garnered significant attention in recent years, beginning in 2008 with Harrow, Hassidim, and Lloyd's quantum algorithm for solving linear systems \cite{harrow2009quantum}.
This burgeoning field has produced exciting quantum algorithms that give hope for finding exponential speedups outside of the now-established gamut of problems related to period finding and Fourier coefficients.
However, in part because of caveats exposited by Aaronson \cite{aaronson2015read}, it is not clear whether any known QML algorithm gives a new exponential speedup over classical algorithms for practically relevant instances of a machine learning problem.
Kerenidis and Prakash's work was notable for addressing all of these caveats, giving a complete quantum algorithm that could be compared directly to classical algorithms.

When Kerenidis and Prakash's work was published, their algorithm was exponentially faster than the best-known classical algorithms.
It was not known whether this was a provably exponential speedup.
The goal of this work is to describe a classical algorithm that performs the same task as the quantum recommendation systems algorithm with only polynomially slower runtime, thus answering this question.
This removes one of the most convincing examples we have of exponential speedups for machine learning problems (see Section~6.7 of Preskill's survey for more context \cite{preskill2018quantum}).

{\bf How the quantum algorithm works.}
Outside of the recommendation systems context, the quantum algorithm just samples from the low-rank approximation of an input matrix.
It proceeds as follows.
First, an application of phase estimation implicitly estimates singular values and locates singular vectors of the input.
A quantum projection procedure then uses this information to project a quantum state with a row of the input to a state with the corresponding row of a low-rank approximation of the input.
Measuring this state samples an entry from the row with probability proportional to its magnitude.
Kerenidis and Prakash posited that, with a classical algorithm, producing samples following these distributions requires time linear in the input dimensions.

The intuition behind this claim is not that singular value estimation or computing projections is particularly difficult computationally.
Rather, it's simply hard to believe that any of these steps can be done without reading the full input.
After all, a significant portion of the theory of low-rank matrix approximation only asks for time complexity linear in input-sparsity or sublinear with some number of passes through the data.
In comparison to these types of results, what the quantum algorithm achieves (query complexity polylogarithmic in the input size) is impressive.

With this claim in mind, Kerenidis and Prakash then apply their quantum algorithm to make a fast online recommendation system.
As information arrives about user-product preferences, we place it into a dynamic data structure that is necessary to run the quantum algorithm.
We can then service requests for recommendations as they arrive by running the quantum algorithm and returning the output sample.
Under strong assumptions about the input data, this sample is likely to be a good recommendation.

{\bf State preparation: the quantum algorithm's assumption.}
To see why the classical algorithm we present is possible, we need to consider the technique Kerenidis and Prakash use to construct their relevant quantum states.

Kerenidis and Prakash's algorithm is one of many QML algorithms \cite{harrow2009quantum,lloyd2013quantum,lloyd2014quantum,kerenidis2017quantum} that require quantum state preparation assumptions, which state that given an input vector $v$, one can quickly form a corresponding quantum state $\ket{v}$.
To achieve the desired runtime in practice, an implementation would replace this assumption with either a procedure to prepare a state from an arbitrary input vector (where the cost of preparation could be amortized over multiple runs of the algorithm) or a specification of input vectors for which quantum state preparation is easy.
Usually QML algorithms abstract away these implementation details, assuming a number of the desired quantum states are already prepared.
The quantum recommendation systems algorithm is unique in that it explicitly comes with a data structure to prepare its states (see Section~\ref{sec:ds}).

These state preparation assumptions are nontrivial: even given ability to query entries of a vector in superposition, preparing states corresponding to arbitrary length-$n$ input vectors is known to take $\Omega(\sqrt{n})$ time (a corollary of quantum search lower bounds \cite{bennett1997strengths}).
Thus, the data structure to quickly prepare quantum states is essential for the recommendation systems algorithm to achieve query complexity polylogarithmic in input size.

{\bf How a classical algorithm can perform as well as the quantum algorithm.}
Our key insight is that the data structure used to satisfy state preparation assumptions can also satisfy $\ell^2$-norm sampling assumptions (defined in Section~\ref{subsec:defsampling}).
So, a classical algorithm whose goal is to ``match'' the quantum algorithm can exploit these assumptions.
The effectiveness of $\ell^2$-norm sampling in machine learning \cite{song2016sublinear,hazan2011beating} and randomized linear algebra \cite{kannan_vempala_2017,doi:10.1137/07070471X} is well-established.
In fact, a work by Frieze, Kannan, and Vempala \cite{frieze2004fast} shows that, with $\ell^2$-norm sampling assumptions, a form of singular value estimation is possible in time independent of $m$ and $n$.
Further, in the context of this literature, sampling from the projection of a vector onto a subspace is not outside the realm of feasibility.
This work just puts these two pieces together.

{\bf The importance of $\ell^2$-norm sampling.}
In an imprecise sense, our algorithm replaces state preparation assumptions with $\ell^2$-norm sampling assumptions.
In this particular case, while quantum superpositions served to represent data implicitly that would take linear time to write out, this need can be served just as well with probability distributions and subsamples of larger pieces of data.

The correspondence between $\ell^2$-norm sampling assumptions and state preparation assumptions makes sense.
While the former sidesteps the obvious search problems inherent in linear algebra tasks by pinpointing portions of vectors or matrices with the most weight, the latter sidesteps these search problems by allowing for quantum states that are implicitly aware of weight distribution.
We suspect that this connection revealed by the state preparation data structure is somewhat deep, and cannot be fixed by simply finding a state preparation data structure without sampling power.

This work demonstrates one major case where classical computing with $\ell^2$-norm sampling is an apt point of comparison for revealing speedups (or, rather, the lack thereof) in QML.
We believe that this reference point remains useful, even for QML algorithms that don't specify state preparation implementation details, and thus are formally incomparable to any classical model.
So, we suggest a general framework for studying the speedups given by QML algorithms with state preparation assumptions: compare QML algorithms with state preparation to classical algorithms with sampling.
Indeed, a QML algorithm using state preparation assumptions {\em should} aim to surpass the capabilities of classical algorithms with $\ell^2$-norm sampling, given that in theory, generic state preparation tends to only appear in settings with generic sampling, and in practice, we already know how to implement fast classical sampling on existing hardware.

In summary, we argue for the following guideline:
{\em when QML algorithms are compared to classical ML algorithms in the context of finding speedups, any state preparation assumptions in the QML model should be matched with $\ell^2$-norm sampling assumptions in the classical ML model.}

\subsection{Recommendation Systems}
In addition to our algorithm having interesting implications for QML, it also can be used as a recommendation system.

To formalize the problem of recommending products to users, we use the following model, first introduced in 1998 by Kumar et al.\ \cite{KUMAR200142} and refined further by Azar et al.\ \cite{azar2001spectral} and Drineas et al.\ \cite{drineas2002competitive}.
We represent the sentiments of $m$ users towards $n$ products with an $m\times n$ {\em preference matrix} $T$, where $T_{ij}$ is large if user $i$ likes product $j$.
We further assume that $T$ is close to a matrix of small rank $k$ (constant or logarithmic in $m$ and $n$), reflecting the intuition that users tend to fall into a small number of classes based on their preferences.
Given a matrix $A$ containing only a subset of entries of $T$, representing our incomplete information about user-product preferences, our goal is to output high-value entries of $T$, representing good recommendations.

Modern work on recommendation systems uses matrix completion to solve this (which works well in practice\footnote{For practical recommendation systems, see Koren et al.\ \cite{koren2009matrix} for a high-level exposition of this technique and Bell et al.\ \cite{bell2007lessons} for more technical details.}), but these techniques must take linear time to produce a recommendation.
Kerenidis and Prakash's recommendation system (and, consequently, this work) follows in an older line of research, which experiments with very strong assumptions on input with the hope of finding a new approach that can drive down runtime to sublinear in $m$ and $n$.
In Kerenidis and Prakash's model, finding a good recommendation for a user $i$ reduces to sampling from the $i$th row of a low rank {\em approximation} of the subsampled data $A$, instead of a low-rank completion.
Using our classical analogue to Kerenidis and Prakash's algorithm, we can get recommendations in $O(\poly(k)\polylog(m,n))$ time, exponentially faster than the best-known in the literature.
Sublinear-time sampling for good recommendations has been proposed before (see introduction of \cite{drineas2002competitive}), but previous attempts to implement it failed to circumvent the bottleneck of needing linear time to write down input-sized vectors.

For context, our model is most similar to the model given in 2002 by Drineas et al.\ \cite{drineas2002competitive}.
However, that algorithm's main goal is minimizing the number of user preferences necessary to generate good recommendations; we discuss in Appendix~\ref{sec:drineas} how to adapt our algorithm to that model to get similar results.
Other approaches include combinatorial techniques \cite{KUMAR200142,Awerbuch:2005:IRS:1070432.1070599} and the use of mixture models \cite{kleinberg2008using}.

We note two major assumptions present in our model that differ from the matrix completion setting.
(The full list of assumptions is given in Section~\ref{sec:model}.)
The first assumption is that our subsample $A$ is contained in a data structure.
This makes sense in the setting of an online recommendation system, where we can amortize the cost of our preprocessing.
Recommendation systems in practice tend to be online, so building a system that keeps data in this data structure to satisfy this assumption seems reasonable.
The second assumption is that we know a constant fraction of the full preference matrix $T$.
This is impractical and worse than matrix completion, for which we can prove correctness given as little as an $\tilde{\Theta}(\frac{k}{m+n})$ fraction of input data \cite{recht2011simpler}.
This seems to be an issue common among works reducing recommendation problems to low-rank matrix approximation \cite{drineas2002competitive,azar2001spectral}.
Nevertheless, we hope that this algorithm, by presenting a novel sampling-based technique with a much faster asymptotic runtime, inspires improved practical techniques and provides an avenue for further research.

\subsection{Algorithm Sketch} \label{subsec:algsketch}
We first state the main result, a classical algorithm that can sample a high-value entry from a given row of a low-rank approximation of a given matrix.
The formal statement can be found in Section~\ref{subsec:mra}.
\begin{mr*}[\ref{main-result-a}, informal]
\mrainformal
\end{mr*}

This makes the runtime {\em independent} of $m$ and $n$.
Here, $(\|A\|_F/\sigma)^2$ is a bound on the rank of the low-rank approximation, so we think of $\sigma$ as something like $\|A\|_F/\sqrt{k}$.
To implement the needed sampling operations, we will use the data structure described in Section~\ref{sec:ds}, which adds at most an additional $O(\log(mn))$ factor in overhead.
This gives a time complexity of
  \[ \tilde{O}\left(\frac{\|A\|_F^{24}}{\sigma^{24}\eps^{12}\eta^6}\log(mn)\frac{\|A_i\|^2}{\|D_i\|^2}\right). \]
This is a large slowdown versus the quantum algorithm in some exponents.
However, we suspect that these exponents can be improved with existing techniques.

The only difference between Theorem~\ref{main-result-a} and its quantum equivalent in \cite{kerenidis2016quantum} is that the quantum algorithm has only logarithmic dependence\footnote{The analysis of the phase estimation in the original paper has some ambiguities, but subsequent work \cite{gilyen2018quantum} demonstrates that essentially the same result can be achieved with a different algorithm.} on $\eps$.
Thus, we can say that our algorithm performs just as well, up to polynomial slowdown and $\eps$ approximation factors.
These $\eps$'s don't affect the classical recommendation system guarantees:

\begin{mr*}[\ref{main-result-b}, informal]
\mrbinformal
\end{mr*}

To prove Theorem~\ref{main-result-a} (Section~\ref{sec:notions}), we present and analyze Algorithm~\ref{alg:full}.
It combines a variety of techniques, all relying on sampling access to relevant input vectors.
The main restriction to keep in mind is that we need to perform linear algebra operations without incurring the cost of reading a full row or column.

The algorithm begins by using the given support for $\ell^2$-norm sampling to run a sampling routine (called \SC{ModFKV}, see Section~\ref{subsec:lra}) based on Frieze, Kannan, and Vempala's 1998 algorithm \cite{frieze2004fast} to find a low-rank approximation of $A$.
It doesn't have enough time to output the matrix in full; instead, it outputs a succinct description of the matrix.
This description is $S$, a normalized constant-sized subset of rows of $A$, along with some constant-sized matrices $\hat{U}$ and $\hat{\Sigma}$, which implicitly describe $\hat{V} := S^T\hat{U}\hat{\Sigma}^{-1}$, a matrix whose columns are approximate right singular vectors of $A$.
The corresponding low-rank approximation is $D := A\hat{V}\hat{V}^T$, an approximate projection of the rows of the input matrix onto the low-dimensional subspace spanned by $\hat{V}$.
Though computing $\hat{V}$ directly takes too much time, we can sample from and query to its columns.
Since rows of $S$ are normalized rows of $A$, we have sampling access to $S$ with our input data structure.
We can translate such samples to samples from $\hat{V}$ using the simple sampling routines discussed in Section~\ref{subsec:vector-samp}.

Though we could use this access to $\hat{V}$ for the rest of our algorithm, we take a more direct approach.
To sample from the $i$th row of $D$ $A_i(S^T\hat{U}\hat{\Sigma}^{-1})(S^T\hat{U}\hat{\Sigma}^{-1})^T$ given $D$'s description, we first estimate $A_iS^T$.
This amounts to estimating a constant number of inner products and can be done with sampling access to $A_i$ by Proposition~\ref{prop:dot-product}.
Then, we multiply this estimate by $\hat{U}\hat{\Sigma}^{-1}(\hat{\Sigma}^{-1})^T\hat{U}^T$, which is a constant-sized matrix.
Finally, we sample from the product of the resulting vector with $S$ and output the result.
This step uses rejection sampling: given the ability to sample and query to a constant-sized set of vectors (in this case, rows of $S$), we can sample from a linear combination of them (Proposition~\ref{prop:rejection-sampling}).

This completes the broad overview of the algorithm.
The correctness and runtime analysis is elementary; most of the work is in showing that \SC{ModFKV}'s various outputs truly behave like approximate large singular vectors and values (Proposition~\ref{prop:modfkv-apporth} and Theorem~\ref{thm:bound-transfer}).

To prove Theorem~\ref{main-result-b} and show that the quality bounds on the recommendations are the same (see Section~\ref{subsec:mrb}), we just follow Kerenidis and Prakash's analysis and apply the model assumptions and theorems (Section~\ref{sec:model}) in a straightforward manner.
The $\ell^2$-norm sampling operations needed to run Algorithm~\ref{alg:full} are instantiated with the data structure Kerenidis and Prakash use (Section~\ref{sec:ds}).

\subsection{Further Questions} \label{subsec:qs}
Since this algorithm is associated both with recommendation systems and quantum machine learning, two lines of questioning naturally follow.

First, we can continue to ask whether any quantum machine learning algorithms have provably exponential speedups over classical algorithms.
We believe that a potentially enlightening approach is to investigate how state preparation assumptions can be satisfied and whether they are in some way comparable to classical sampling assumptions.
After all, we find it unlikely that a quantum exponential speedup can be reinstated just with a better state preparation data structure.
However, we are unaware of any research in this area in particular, which could formalize a possible connection between QML algorithms with state preparation assumptions and classical ML algorithms with sampling assumptions.

Second, while the recommendation system algorithm we give is asymptotically exponentially faster than previous algorithms, there are several aspects of this algorithm that make direct application infeasible in practice.
First, the model assumptions are somewhat constrictive.
It is unclear whether the algorithm still performs well when such assumptions are not satisfied.
Second, the exponents and constant factors are large (mostly as a result of using Frieze, Kannan, and Vempala's algorithm \cite{frieze2004fast}).
We believe that the ``true'' exponents are much smaller and could result from more sophisticated techniques (see, for example, \cite{deshpande2006adaptive}).

\section{Definitions} \label{sec:definitions}
Throughout, we obey the following conventions.
We assume that basic operations on input data (e.g.\ adding, multiplying, reading, and writing) take $O(1)$ time.
$[n] := \{1,\ldots,n\}$.
$f \lesssim g$ denotes the ordering $f = O(g)$ (and correspondingly for $\gtrsim$ and $\eqsim$).
For a matrix $A$, $A_i$ and $A^{(i)}$ will refer to its $i$th row and column, respectively.
$\|A\|_F$ and $\|A\|_2$ will refer to Frobenius and spectral norm, respectively.
Norm of a vector $v$, denoted $\|v\|$, will always refer to $\ell^2$-norm.
The absolute value of $x \in \BB{R}$ will be denoted $|x|$.
Occasionally, matrix and vector inequalities of the form $\|x-y\| \leq \eps$ will be phrased in the form $x = y+E$, where $\|E\| \leq \eps$.
Thus, the letter $E$ will always refer to some form of perturbation or error.

For a matrix $A \in \BB{R}^{m\times n}$, let $A = U\Sigma V^T = \sum_{i=1}^{\min m,n} \sigma_i u_iv_i^T$ be the SVD of $A$.
Here, $U \in \BB{R}^{m\times m}$ and $V \in \BB{R}^{n\times n}$ are unitary matrices with columns $\{u_i\}_{i \in [m]}$ and $\{v_i\}_{i \in [n]}$, the left and right singular vectors, respectively.
$\Sigma \in \BB{R}^{m\times n}$ is diagonal with $\sigma_i := \Sigma_{ii}$ and the $\sigma_i$ nonincreasing and nonnegative.

We will use the function $\ell$ to indicate splitting the singular vectors along a singular value:
  \[ \ell(\lambda) := \max \{i \mid \sigma_i \geq \lambda\}.\]
For example, $\sigma_1$ through $\sigma_{\ell(\lambda)}$ gives all of the singular values that are at least $\lambda$.
This notation suppresses $\ell$'s dependence on $\sigma_i$, but it will always be clear from context.

$\Pi$ will always refer to an orthogonal projector.
That is, if $\beta = \{b_1,\ldots,b_d\}$ is an orthonormal basis for $\operatorname{im}\Pi$, then $\Pi = \sum_{i=1}^d b_ib_i^T = BB^T$ for $B$ the matrix whose columns are the elements of $\beta$.
We will often conflate $B$, the matrix of basis vectors, and the basis $\beta$ itself.

\subsection{Low-Rank Approximations} \label{subsec:lr-notation}
We will use various techniques to describe low-rank approximations of $A$.
All of these techniques will involve projecting the rows onto some span of right singular vectors.
\begin{align*}
  A_k &:= A\Pi_k &
  \operatorname{im}\Pi_k &:= \operatorname{span}\{v_i \mid i \in [k]\} \\
  A_{\sigma} &:= A\Pi_{\sigma} &
  \operatorname{im}\Pi_{\sigma} &:= \operatorname{span}\{v_i \mid i \in [\ell(\sigma)]\}
\end{align*}
$A_k$ and $A_{\sigma}$ correspond to the standard notions of low-rank approximations of $A$.
Thus, $A_k = \sum_{i=1}^k \sigma_iu_iv_i^T$ and is a rank-$k$ matrix minimizing the Frobenius norm distance from $A$.
Similarly, $A_{\sigma}$ is just $A_t$ for $t = \ell(\sigma)$.
Notice that $\operatorname{rank} A_{ \frac{\|A\|_F}{\sqrt{\lambda}}} \leq \lambda$.

We will need to relax this notion for our purposes, and introduce error $\eta \in [0,1]$.
Define $A_{\sigma,\eta} := AP_{\sigma,\eta}$ where $P_{\sigma,\eta}$ is some Hermitian matrix satisfying $\Pi_{\sigma(1+\eta)} \preceq P_{\sigma,\eta} \preceq \Pi_{\sigma(1-\eta)}$ and $\preceq$ is the Loewner order.
In words, $A_{\sigma,\eta}$ is the class of matrices ``between'' $A_{\sigma(1+\eta)}$ and $A_{\sigma(1-\eta)}$: $P_{\sigma,\eta}$ is the identity on $v_i$'s with $i \leq \ell(\sigma(1+\eta))$, the zero map on $v_i$'s with $i > \ell(\sigma(1-\eta))$, and some PSD matrix with norm at most on the subspace spanned by $v_i$'s with $i \in (\ell(\sigma(1+\eta)), \ell(\sigma(1-\eta))]$.
Such a form of error could arise from having $\eta$-like error in estimating the singular values used to compute a low-rank matrix approximation.
$\eta$ should be thought of as constant ($1/5$ will be the eventual value), and $\sigma$ should be thought of as very large (say, a constant multiple of $\|A\|_F$), so $A_{\sigma,\eta}$ always has low rank.

\subsection{Sampling} \label{subsec:defsampling}
For a nonzero vector $x \in \BB{R}^n$, we denote by $\CC{D}_x$ the distribution over $[n]$ whose probability density function is
  \[ \CC{D}_x(i) = \frac{x_i^2}{\|x\|^2} \]
We will call a sample from $\CC{D}_x$ a sample from $x$.

We make two basic observations.
First, $\CC{D}_x$ is the distribution resulting from measuring the quantum state $\ket{x} := \frac{1}{\|x\|}\sum x_i\ket{i}$ in the computational basis.
Second, sampling access to $\CC{D}_x$ makes easy some tasks that are hard given just query access to $x$.
For example, while finding a hidden large entry of $x \in \BB{R}^n$ takes $\Omega(n)$ queries with just query access, it takes a constant number of samples with query and sample access.

In all situations, sampling access will be present alongside query access, and accordingly, we will conflate samples $i \sim \CC{D}_x$ with the corresponding entries $x_i$.
Note that knowledge of $\|x\|$ is also relevant and useful in this sampling context, since it allows for computing probabilities from $\CC{D}_x$ and yet is hard to compute even with query and sampling access to $x$.

For probability distributions $P,Q$ (as density functions) over a (discrete) universe $X$, the total variation distance between them is defined as
  \[ \|P - Q\|_{TV} := \frac{1}{2}\sum_{x \in X} \Big|P(x) - Q(x)\Big|. \]
For a set $S$, we denote pulling an $s \in S$ uniformly at random by $s \sim_u S$.
We will continue to conflate a distribution with its density function.

\section{Data Structure} \label{sec:ds}
Since we are interested in achieving sublinear bounds for our algorithm, we need to concern ourselves with how the input is given.

In the recommendation systems context, entries correspond to user-product interactions, so we might expect that the input matrix $A \in \BB{R}^{m\times n}$ is given as an unordered stream of entries $(i, j, A_{ij})$.
However, if the entries are given in such an unprocessed format, then clearly linear time is required even to parse the input into a usable form.
Even when the input is relatively structured (for example, if we are given the known entries of $T$ sorted by row and column), there is no hope to sample the low-rank approximation of a generic matrix in sublinear time because of the time needed to locate a nonzero entry.

To avoid these issues, we will instead consider our input matrix stored in a low-overhead data structure.
We define it first for a vector, then for a matrix.

\begin{lemma} \label{lem:ds-vector}
  There exists a data structure storing a vector $v \in \BB{R}^n$ with $w$ nonzero entries in $O(w\log (n))$ space, supporting the following operations:
  \begin{itemize}
    \item Reading and updating an entry of $v$ in $O(\log n)$ time;
    \item Finding $\|v\|^2$ in $O(1)$ time;
    \item Sampling from $\CC{D}_v$ in $O(\log n)$ time.
  \end{itemize}
\end{lemma}

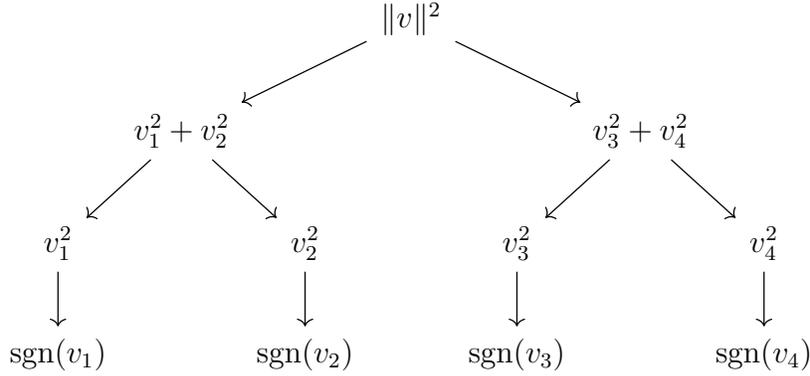
\begin{figure}[ht]
\centering
\[ \begin{tikzcd}[column sep=0]
  & & & \|v\|^2 \ar[dll] \ar[drr] & & & \\
  & v_1^2 + v_2^2 \ar[dl] \ar[dr] & & & & v_3^2 + v_4^2 \ar[dl] \ar[dr] & \\
  v_1^2 \ar[d] & & v_2^2 \ar[d] & & v_3^2 \ar[d] & & v_4^2 \ar[d] \\
  \operatorname{sgn}(v_1) & & \operatorname{sgn}(v_2) & & \operatorname{sgn}(v_3) & & \operatorname{sgn}(v_4)
\end{tikzcd} \]
\caption{Binary search tree (BST) data structure for $v \in \BB{R}^4$.
The leaf nodes store $v_i$ via its weight $v_i^2$ and sign $\operatorname{sgn}(v_i)$, and the weight of an interior node is just the sum of the weights of its children.
To update an entry, update all of the nodes above the corresponding leaf.
To sample from $\CC{D}_v$, start from the top of the tree and randomly recurse on a child with probability proportional to its weight.
To take advantage of sparsity, prune the tree to only nonzero nodes.
}
\end{figure}

\begin{proposition} \label{prop:ds-matrix}
  Consider a matrix $A \in \BB{R}^{m\times n}$.
  Let $\tilde{A} \in \BB{R}^m$ be a vector whose $i$th entry is $\|A_i\|$.
  There exists a data structure storing a matrix $A \in \BB{R}^{m\times n}$ with $w$ nonzero entries in $O(w\log mn)$ space, supporting the following operations:
  \begin{itemize}
    \item Reading and updating an entry of $A$ in $O(\log mn)$ time;
    \item Finding $\tilde{A}_i$ in $O(\log m)$ time;
    \item Finding $\|A\|_F^2$ in $O(1)$ time;
    \item Sampling from $\CC{D}_{\tilde{A}}$ and $\CC{D}_{A_i}$ in $O(\log mn)$ time.
  \end{itemize}
\end{proposition}
This can be done by having a copy of the data structure specified by Lemma~\ref{lem:ds-vector} for each row of $A$ and $\tilde{A}$ (which we can think of as the roots of the BSTs for $A$'s rows).
This has all of the desired properties, and in fact, is the data structure Kerenidis and Prakash use to prepare arbitrary quantum states (Theorem~A.1 in \cite{kerenidis2016quantum}).
Thus, our algorithm can operate on the same input, although any data structure supporting the operations detailed in Proposition~\ref{prop:ds-matrix} will also suffice.

This data structure and its operations are not as ad hoc as they might appear.
The operations listed above appear in other work as an effective way to endow a matrix with $\ell^2$-norm sampling assumptions \cite{drineas2002competitive,frieze2004fast}.

\section{Main Algorithm} \label{sec:notions}

Our goal is to prove Theorem~\ref{main-result-a}:

\begin{theorem*}[\ref{main-result-a}]
  \mra
\end{theorem*}

We present the algorithm (Algorithm~\ref{alg:full}) and analysis nonlinearly.
First, we give two algorithms that use $\ell^2$-norm sampling access to their input vectors to perform basic linear algebra.
Second, we present \SC{ModFKV}, a sampling algorithm to find the description of a low-rank matrix approximation.
Third, we use the tools we develop to get from this description to the desired sample.

\subsection{Vector Sampling} \label{subsec:vector-samp}

Recall how we defined sampling from a vector.
\begin{definition}
  For a vector $x \in \BB{R}^n$, we denote by $\CC{D}_x$ the distribution over $[n]$ with density function $\CC{D}_x(i) = x_i^2/\|x\|^2$.
  We call a sample from $\CC{D}_x$ a sample from $x$.
\end{definition}

We will need that closeness of vectors in $\ell^2$-norm implies closeness of their respective distributions in TV distance:
\begin{lemma} \label{lem:2norm-to-TV}
  For $x,y \in \BB{R}^n$ satisfying $\|x-y\| \leq \eps$, the corresponding distributions $\CC{D}_x$, $\CC{D}_y$ satisfy $\|\CC{D}_x-\CC{D}_y\|_{TV} \leq 2\eps/\|x\|$.
\end{lemma}
\begin{proof}
  Let $\hat{x}$ and $\hat{y}$ be the normalized vectors $x/\|x\|$ and $y/\|y\|$.
  \begin{multline*}
    \|\CC{D}_x-\CC{D}_y\|_{TV} = \frac{1}{2}\sum_{i=1}^n \big|\hat{x}_i^2 - \hat{y}_i^2\big|
    = \frac{1}{2}\big\langle|\hat{x}-\hat{y}|,|\hat{x}+\hat{y}|\big\rangle
    \leq \frac{1}{2}\|\hat{x}-\hat{y}\|\|\hat{x}+\hat{y}\| \\
    \leq \|\hat{x}-\hat{y}\|
    = \frac{1}{\|x\|}\Big\|x-y-(\|x\|-\|y\|)\hat{y}\Big\|
    \leq \frac{1}{\|x\|}\Big(\|x-y\|+\big|\|x\|-\|y\|\big|\Big)
    \leq \frac{2\eps}{\|x\|}
  \end{multline*}
  The first inequality follows from Cauchy-Schwarz, and the rest follow from triangle inequality.
\end{proof}

Now, we give two subroutines that can be performed, assuming some vector sampling access.
First, we show that we can estimate the inner product of two vectors well.
\begin{proposition} \label{prop:dot-product}
  Given query access to $x,y \in \BB{R}^n$, sample access to $\CC{D}_x$, and knowledge of $\|x\|$, $\langle x,y\rangle$ can be estimated to additive error $\|x\|\|y\|\eps$ with at least $1-\delta$ probability using $O(\frac{1}{\eps^2}\log\frac{1}{\delta})$ queries and samples (and the same time complexity).
\end{proposition}
\begin{proof}
  Perform samples in the following way: for each $i$, let the random variable $Z$ be $y_i/x_i$ with probability $x_i^2/\|x\|^2$ (so we sample $i$ from $\CC{D}_x$).
  We then have:
  \begin{align*}
    \operatorname{E}[Z] &= \sum \frac{y_i}{x_i}\frac{x_i^2}{\|x\|^2} = \frac{\sum x_iy_i}{\|x\|^2} = \frac{\langle x,y\rangle}{\|x\|^2}, \\
    \operatorname{Var}[Z] &\leq \sum \left(\frac{y_i}{x_i}\right)^2\frac{x_i^2}{\|x\|^2} = \frac{\sum y_i^2}{\|x\|^2} = \frac{\|y\|^2}{\|x\|^2}.
  \end{align*}
  Since we know $\|x\|$, we can normalize by it to get a random variable whose mean is $\langle x,y\rangle$ and whose standard deviation is $\sigma = \|x\|\|y\|$.

  The rest follows from standard techniques: we take the median of $6\log \frac{1}{\delta}$ copies of the mean of $\frac{9}{2\eps^2}$ copies of $Z$ to get within $\eps\sigma = \eps\|x\|\|y\|$ of $\langle x,y\rangle$ with probability at least $1-\delta$ in $O(\frac{1}{\eps^2}\log\frac{1}{\delta})$ accesses.
  All of the techniques used here take linear time.
\end{proof}

Second, we show that, given sample access to some vectors, we can sample from a linear combination of them.

\begin{proposition} \label{prop:rejection-sampling}
  Suppose we are given query and sample access to the columns of $V \in \BB{R}^{n\times k}$, along with their norms.
  Then given $w \in \BB{R}^k$ (as input), we can output a sample from $Vw$ in $O(k^2C(V,w))$ expected query complexity and expected time complexity, where
    \[ C(V,w) := \frac{\sum_{i=1}^k\|w_iV^{(i)}\|^2}{\|Vw\|^2}. \]
\end{proposition}
$C$ measures the amount of cancellation for $Vw$.
For example, when the columns of $V$ are orthogonal, $C=1$ for all nonzero $w$, since there is no cancellation.
Conversely, when the columns of $V$ are linearly dependent, there is a choice of nonzero $w$ such that $\|Vw\| = 0$, maximizing cancellation.
In this context, $C$ is undefined, which matches with sampling from the zero vector also being undefined.
By perturbing $w$ we can find vectors requiring arbitrarily large values of $C$.
\begin{proof}
  We use rejection sampling: see Algorithm~\ref{alg:rejsamp}.
  Given sampling access to a distribution $P$, rejection sampling allows for sampling from a ``close'' distribution $Q$, provided we can compute some information about their corresponding distributions.

  \begin{algorithm} 
    \caption{Rejection Sampling}
    \SetAlgoLined \label{alg:rejsamp}
    Pull a sample $s$ from $P$\;
    Compute $r_s = \frac{Q(s)}{MP(s)}$ for some constant $M$\;
    Output $s$ with probability $r_s$ and restart otherwise\;
  \end{algorithm}

  If $r_i \leq 1$ for all $i$, then the above procedure is well-defined and outputs a sample from $Q$ in $M$ iterations in expectation.\footnote{The number of iterations is a geometric random variable, so this can be converted into a bound guaranteeing a sample in $M\log 1/\delta$ iterations with failure probability $1-\delta$, provided the algorithm knows $M$. All expected complexity bounds we deal with can be converted to high probability bounds in the manner described.}

  In our case, $P$ is the distribution formed by first sampling a row $j$ with probability proportional to $\|w_jV^{(j)}\|^2$ and then sampling from $\CC{D}_{V^{(j)}}$; $Q$ is the target $\CC{D}_{Vw}$. We choose
    \[ r_i = \frac{(Vw)_i^2}{k \sum_{j=1}^k (V_{ij}w_j)^2}, \]
  which we can compute in $k$ queries\footnote{Notice that we can compute $r_i$ without directly computing the probabilities $Q(i)$. This helps us because computing $Q(i)$ involves computing $\|Vw\|$, which is nontrivial.}.
  This expression is written in a way the algorithm can directly compute, but it can be put in the form of the rejection sampling procedures stated above:
    \[ M = \frac{Q(i)k \sum_{j=1}^k (V_{ij}w_j)^2}{P(i)(Vw)_i^2}
    = \frac{k(\sum_{j=1}^k\|w_jV^{(j)}\|^2)}{\|Vw\|^2}
    = kC(V,w). \]
  $M$ is independent of $i$, so it is a constant as desired.
  To prove correctness, all we need to show is that our choice of $r_i$ is always at most $1$.
  This follows from Cauchy-Schwarz:
    \[ r_i = \frac{(Vw)_i^2}{k \sum_{j=1}^k (V_{ij}w_j)^2} = \frac{(\sum_{j=1}^k V_{ij}w_j)^2}{k \sum_{j=1}^k (V_{ij}w_j)^2} \leq 1. \]
  Each iteration of the procedure takes $O(k)$ queries, leading to a query complexity of $O(k^2C(V,w))$.
  Time complexity is linear in the number of queries.
\end{proof}

\subsection{Finding a Low-Rank Approximation} \label{subsec:lra}

Now, we describe the low-rank approximation algorithm that we use at the start of the main algorithm.

\begin{theorem} \label{thm:modfkv-main}
  Given a matrix $A \in \BB{R}^{m\times n}$ supporting the sample and query operations described in Proposition~\ref{prop:ds-matrix}, along with parameters $\sigma \in (0,\|A\|_F], \eps \in (0,\sqrt{\sigma/\|A\|_F}/4], \eta \in [\eps^2,1]$, there is an algorithm that outputs a succinct description (of the form described below) of some $D$ satisfying $\|D-A_{\sigma,\eta}\|_F \leq \eps\|A\|_F$ with probability at least $1-\delta$ and
    \[ O\left(\poly\Big(\frac{\|A\|_F^2}{\sigma^2},\frac{1}{\eps},\frac{1}{\eta},\log \frac{1}{\delta}\Big)\right) \]
  query and time complexity.
\end{theorem}

To prove this theorem, we modify the algorithm given by Frieze, Kannan, and Vempala \cite{frieze2004fast} and show that it satisfies the desired properties.
The modifications are not crucial to the correctness of the full algorithm: without them, we simply get a different type of low-rank approximation bound.
They come into play in Section~\ref{sec:model} when proving guarantees about the algorithm as a recommendation system.

\begin{algorithm} 
  \caption{\SC{ModFKV}}
  \KwIn{Matrix $A \in \BB{R}^{m\times n}$ supporting operations in Proposition~\ref{prop:ds-matrix}, threshold $\sigma$, error parameters $\eps,\eta$}
  \KwOut{A description of an output matrix $D$}
  \SetAlgoLined \label{alg:ModFKV}
    Set $K = \|A\|_F^2/\sigma^2$ and $\bar{\eps} = \eta\eps^2$\;
    Set $q = \Theta\big(\frac{K^4}{\bar{\eps}^2}\big)$\;
    Sample rows $i_1,\ldots,i_q$ from $\CC{D}_{\tilde{A}}$\;
    Let $\CC{F}$ denote the distribution given by choosing an $s \sim_u [q]$, and choosing a column from $\CC{D}_{A_{i_s}}$\;
    Sample columns $j_1,\ldots,j_q$ from $\CC{F}$\;
    Let $W$ be the resulting $q\times q$ row-and-column-normalized submatrix
      $W_{rc} := \frac{A_{i_rj_c}}{q\sqrt{\CC{D}_{\tilde{A}}(i_r)\CC{F}(j_c)}}$\;
    Compute the left singular vectors of $W$ $u^{(1)},\ldots,u^{(k)}$ that correspond to singular values $\sigma^{(1)},\ldots,\sigma^{(k)}$ larger than $\sigma$\;
    Output $i_1,\ldots,i_q$, $\hat{U} \in \BB{R}^{q\times k}$ the matrix whose $i$th column is $u^{(i)}$, and $\hat{\Sigma} \in \BB{R}^{k\times k}$ the diagonal matrix whose $i$th entry on the diagonal is $\sigma^{(i)}$.
    This is the description of the output matrix $D$\;
\end{algorithm}

The algorithm, \SC{ModFKV}, is given in Algorithm~\ref{alg:ModFKV}.
It subsamples the input matrix, computes the subsample's large singular vectors and values, and outputs them with the promise that they give a good description of the singular vectors of the full matrix.
We present the algorithm as the original work does, aiming for a constant failure probability.
This can be amplified to $\delta$ failure probability by increasing $q$ by a factor of $O(\log\frac1\delta)$ (the proof uses a martingale inequality; see Theorem~1 of \cite{doi:10.1137/S0097539704442684}).
More of the underpinnings are explained in Frieze, Kannan, and Vempala's paper \cite{frieze2004fast}.

We get the output matrix $D$ from its description in the following way.
Let $S$ be the submatrix given by restricting the rows to $i_1,\ldots,i_q$ and renormalizing row $i$ by $1/\sqrt{q\CC{D}_{\tilde{A}}(i)}$ (so they all have the same norm).
Then $\hat{V} := S^T\hat{U}\hat{\Sigma}^{-1} \in \BB{R}^{n\times k}$ is our approximation to the large right singular vectors of $A$; this makes sense if we think of $S$, $\hat{U}$, and $\hat{\Sigma}$ as our subsampled low-rank approximations of $A$, $U$, and $\Sigma$ (from $A$'s SVD).
Appropriately, $D$ is the ``projection'' of $A$ onto the span of $\hat{V}$:
    \[ D := A\hat{V}\hat{V}^T = AS^T\hat{U}\hat{\Sigma}^{-2}\hat{U}^TS. \]

The query complexity of $\SC{ModFKV}$ is dominated by querying all of the entries of $W$, which is $O(q^2)$, and the time complexity is dominated by computing $W$'s SVD, which is $O(q^3)$.
We can convert this to the input parameters using that $q = O(\frac{\|A\|^8}{\sigma^8\eps^4\eta^2})$.

\SC{ModFKV} differs from \SC{FKV} only in that $\sigma$ is taken as input instead of $k$, and is used as the threshold for the singular vectors.
As a result of this change, $K$ replaces $k$ in the subsampling steps, and $\sigma$ replaces $k$ in the SVD step.
Notice that the number of singular vectors taken (denoted $k$) is at most $K$, so in effect, we are running \SC{FKV} and just throwing away some of the smaller singular vectors.
Because we ignore small singular values that \SC{FKV} had to work to find, we can sample a smaller submatrix, speeding up our algorithm while still achieving an analogous low-rank approximation bound:
\begin{lemma} \label{lem:fkvbound}
  The following bounds hold for the output matrix $D$ (here, $k$ is the width of $\hat{V}$, and thus a bound on $\rank D$):
  \begin{gather*} 
    \|A-D\|_F^2 \leq \|A-A_k\|_F^2 + \bar{\eps}\|A\|_F^2 \label{eqn:fkvbound}\tag{$\diamondsuit$}\\
    \text{and }\ell((1+\bar{\eps}\sqrt{K})\sigma) \leq k \leq \ell((1-\bar{\eps}\sqrt{K})\sigma). \label{eqn:kbound} \tag{$\heartsuit$}
  \end{gather*}
\end{lemma}

The following property will be needed to prove correctness of $\SC{ModFKV}$ and Algorithm~\ref{alg:full}.
The estimated singular vectors in $\hat{V}$ behave like singular vectors, in that they are close to orthonormal.
\begin{proposition} \label{prop:modfkv-apporth}
  The output vectors $\hat{V}$ satisfy
    \[ \|\hat{V} - \Lambda\|_F = O(\bar{\eps}) \]
  for $\Lambda$ a set of orthonormal vectors with the same image as $\hat{V}$.
\end{proposition}
As an easy corollary, $\hat{V}\hat{V}^T$ is $O(\bar{\eps})$-close in Frobenius norm to the projector $\Lambda\Lambda^T$, since $\hat{V}\hat{V}^T = (\Lambda+E)(\Lambda+E)^T$ and $\|\Lambda E^T\|_F = \|E\Lambda^T\|_F = \|E\|_F$.
The proofs of the above lemma and proposition delve into FKV's analysis, so we defer them to the appendix.

The guarantee on our output matrix $D$ is (\ref{eqn:fkvbound}), but for our recommendation system, we want that $D$ is close to some $A_{\sigma,\eta}$.
Now, we present the core theorem showing that the former kind of error implies the latter.
\begin{theorem} \label{thm:bound-transfer}
  If $\Pi$ a $k$-dimensional orthogonal projector satisfies
    \[ \|A_k\|_F^2 \leq \|A\Pi\|_F^2 + \eps\sigma_k^2, \]
  then
    \[ \|A\Pi - A_{\sigma_k,\eta}\|_F^2 \lesssim \eps\sigma_k^2/\eta, \]
  where $\eps \leq \eta \leq 1$.\footnote{An analogous proof gives the more general bound $\|\Pi - P_{\sigma,\eta}\|_F^2 \lesssim \eps/\eta$.}
\end{theorem}
The proof is somewhat involved, so we defer it to the appendix.
To our knowledge, this is a novel translation of a typical FKV-type bound as in (\ref{eqn:fkvbound}) to a new, useful type of bound, so we believe this theorem may find use elsewhere.
Now, we use this theorem to show that $D$ is close to some $A_{\sigma,\eta}$.
\begin{corollary} \label{cor:modfkv-good}
  $\|D - A_{\sigma,\eta}\|_F \lesssim \eps\|A\|_F/\sqrt{\eta}$.
\end{corollary}
\begin{proof}
  Throughout the course of this proof, we simplify and apply theorems based on the restrictions on the parameters in Theorem~\ref{thm:modfkv-main}.

  First, notice that the bound (\ref{eqn:fkvbound}) can be translated to the type of bound in the premise of Theorem~\ref{thm:bound-transfer}, using Proposition~\ref{prop:modfkv-apporth}.
  \begin{align*}
    \|A-D\|_F^2 &\leq \|A-A_k\|_F^2 + \bar{\eps}\|A\|_F^2 \\
    \|A-A(\Lambda\Lambda^T+E)\|_F^2 &\leq \|A-A_k\|_F^2 + \bar{\eps}\|A\|_F^2 \\
    (\|A-A\Lambda\Lambda^T\|_F-\bar{\eps}\|A\|_F)^2 &\lesssim \|A-A_k\|_F^2 + \bar{\eps}\|A\|_F^2 \\
    \|A\|_F^2-\|A\Lambda\Lambda^T\|_F^2 &\lesssim \|A\|_F^2-\|A_k\|_F^2 + \bar{\eps}\|A\|_F^2 \\
    \|A_k\|_F^2 &\lesssim \|A\Lambda\Lambda^T\|_F^2 + (\bar{\eps}\|A\|_F^2/\sigma_k^2)\sigma_k^2
  \intertext{The result of the theorem is that}
    \Big\|A\Lambda\Lambda^T - A_{\sigma_k,\frac{\eta-\bar{\eps}\sqrt{K}}{1-\bar{\eps}\sqrt{K}}}\Big\|_F^2 &\lesssim \frac{(\bar{\eps}\|A\|_F^2/\sigma_k^2)\sigma_k^2}{\frac{\eta-\bar{\eps}\sqrt{K}}{1-\bar{\eps}\sqrt{K}}}
    \lesssim \frac{\bar{\eps}}{\eta}\|A\|_F^2.
  \end{align*}
  The bound on $k$ (\ref{eqn:kbound}) implies that any $A_{\sigma_k,\frac{\eta-\bar{\eps}\sqrt{K}}{1-\bar{\eps}\sqrt{K}}}$ is also an $A_{\sigma,\eta}$ (the error of the former is contained in the latter), so we can conclude
    \[ \|D - A_{\sigma,\eta}\|_F \lesssim \|A\Lambda\Lambda^T - A_{\sigma,\eta}\|_F + \bar{\eps}\|A\|_F \lesssim \sqrt{\frac{\bar{\eps}}{\eta}}\|A\|_F. \]
  $\bar{\eps}$ was chosen so that the final term is bounded by $\eps\|A\|_F$.
\end{proof}

This completes the proof of Theorem~\ref{thm:modfkv-main}.

To summarize, after this algorithm we are left with the description of our low-rank approximation $D = AS^T\hat{U}\Sigma^{-2}\hat{U}^TS$, which will suffice to generate samples from rows of $D$.
It consists of the following:
\begin{itemize}
  \item $\hat{U} \in \BB{R}^{q \times k}$, explicit orthonormal vectors;
  \item $\hat{\Sigma} \in \BB{R}^{k\times k}$, an explicit diagonal matrix whose diagonal entries are in $(\sigma,\|A\|_F]$;
  \item $S \in \BB{R}^{q \times n}$, which is {\em not} output explicitly, but whose rows are rows of $A$ normalized to equal norm $\|A\|_F/\sqrt{q}$ (so we can sample from $S$'s rows); and
  \item $\hat{V} \in \BB{R}^{n\times k}$, a close-to-orthonormal set of vectors implicitly given as $S^T\hat{U}\hat{\Sigma}^{-1}$.
\end{itemize}

\subsection{Proof of Theorem 1} \label{subsec:mra}
\setcounter{mr}{0}
\begin{mr} \label{main-result-a}
  \mra
\end{mr}
\begin{proof}
We will give an algorithm (Algorithm~\ref{alg:full}) where the error in the output distribution is $O(\eps\|A_i\|/\|D_i\|)$-close to $\CC{D}_{D_i}$, and there is no dependence on $\|A_i\|/\|D_i\|$ in the runtime, and discuss later how to modify the algorithm to get the result in the theorem.

\begin{algorithm}
  \caption{Low-rank approximation sampling}
  \KwIn{Matrix $A \in \BB{R}^{m\times n}$ supporting the operations in \ref{prop:ds-matrix}, user $i \in [m]$, threshold $\sigma$, $\eps > 0$, $\eta \in (0,1]$}
  \KwOut{Sample $s \in [n]$} \label{alg:full}
  Run \SC{ModFKV} (\ref{alg:ModFKV}) with parameters $(\sigma,\eps,\eta)$ to get a description of $D = A\hat{V}\hat{V}^T = AS^T\hat{U}\hat{\Sigma}^{-2}\hat{U}^TS$\;
  Estimate $A_iS^T$ entrywise by using Proposition~\ref{prop:dot-product} with parameter $\frac{\eps}{\sqrt{K}}$ to estimate $\langle A_i,S^T_t\rangle$ for all $t \in [q]$. Let $\est$ be the resulting $1\times q$ vector of estimates\;
  Compute $\est \hat{U}\hat{\Sigma}^{-2}\hat{U}^T$ with matrix-vector multiplication\;
  Sample $s$ from $(\est\hat{U}\hat{\Sigma}^{-2}\hat{U}^T)S$ using Proposition~\ref{prop:rejection-sampling}\;
  Output $s$\;
  \SetAlgoLined
\end{algorithm}

{\bf Correctness:}
By Theorem~\ref{thm:modfkv-main}, for sufficiently small\footnote{This is not a strong restriction: $\eps \lesssim \min\{\sqrt{\eta},\sqrt{\sigma/\|A\|_F}\}$ works. This makes sense: for $\eps$ any larger, the error can encompass addition or omission of full singular vectors.} $\eps$, the output matrix $D$ satisfies
    \[ \|D-A_{\sigma,\eta}\|_F \leq \eps\|A\|_F. \]
So, all we need is to approximately sample from the $i$th row of $D$, given its description.

Recall that the rows of $S_t^T$ have norm $\|A\|_F/\sqrt{q}$. Thus, the guarantee from Proposition~\ref{prop:dot-product} states that each estimate of an entry has error at most $\frac{\eps}{\sqrt{Kq}}\|A_i\|\|A\|_F$, meaning that $\|\est - A_iS^T\| \leq \frac{\eps}{\sqrt{K}}\|A_i\|\|A\|_F$.
Further, using that $\hat{V}$ is close to orthonormal (Proposition~\ref{prop:modfkv-apporth}) and $\|\hat{U}\Sigma^{-1}\| \leq \frac{1}{\sigma}$, we have that the vector we sample from is close to $D_i$:
\begin{multline*}
    \|(\est - A_iS^T)\hat{U}\Sigma^{-1}\hat{V}^T\|
    \leq (1+O(\eps^2))\|\est\hat{U}\Sigma^{-1} - A_iS^T\hat{U}\Sigma^{-1}\| \\
    \lesssim \frac{1}{\sigma}\|\est - A_iS^T\|
    \leq \frac{\eps}{\sigma\sqrt{K}}\|A_i\|\|A\|_F = \eps\|A_i\|
\end{multline*}
Finally, by Lemma~\ref{lem:2norm-to-TV}, we get the desired bound: that the distance from the output distribution to $\CC{D}_{D_i}$ is $O(\eps\|A_i\|/\|D_i\|)$.

{\bf Runtime:} Applying Proposition~\ref{prop:dot-product} $q$ times takes $O(\frac{Kq}{\eps^2}\log\frac q\delta)$ time; the naive matrix-vector multiplication takes $O(Kq)$ time; and applying Proposition~\ref{prop:rejection-sampling} takes time $O(Kq^2)$, since
\begin{multline*}
  C(S^T, \hat{U}\hat{\Sigma}^{-2}\hat{U}^T\est^T)
  = \frac{\sum_{j=1}^q \|(\est\hat{U}\hat{\Sigma}^{-2}\hat{U}^T)_jS_j\|^2}{\|\est\hat{U}\hat{\Sigma}^{-2}\hat{U}^TS\|^2}
  \leq \frac{\|\est\hat{U}\hat{\Sigma}^{-2}\hat{U}^T\|^2\|S\|_F^2}{\|\est\hat{U}\hat{\Sigma}^{-2}\hat{U}^TS\|^2} \\
  \leq \frac{\|\hat{\Sigma}^{-1}\hat{U}^T\|^2\|S\|_F^2}{\min_{x:\|x\|=1}\|x\Sigma^{-1}\hat{U}^TS\|^2}
  \lesssim \frac{\|A\|_F^2}{\sigma^2(1-\eps^2)^2} = O(K)
\end{multline*}
using Cauchy-Schwarz, Proposition~\ref{prop:modfkv-apporth}, and the basic facts about $D$'s description\footnote{We have just proved that, given $D$'s description, we can sample from any vector of the form $\hat{V}x$ in $O(Kq^2)$ time.}.

The query complexity is dominated by the use of Proposition~\ref{prop:rejection-sampling} and the time complexity is dominated by the $O(q^3)$ SVD computation in \SC{ModFKV}, giving
\begin{gather*}
  \text{Query complexity} = \tilde{O}\left(\frac{\|A\|_F^2}{\sigma^2}\Big(\frac{\|A\|_F^8}{\sigma^8\eps^4\eta^2}\Big)^2\right) = \tilde{O}\left(\frac{\|A\|_F^{18}}{\sigma^{18}\eps^8\eta^4}\right) \\
  \text{Time complexity} = \tilde{O}\left(\frac{\|A\|_F^{24}}{\sigma^{24}\eps^{12}\eta^6}\right),
\end{gather*}
where the $\tilde{O}$ hides the $\log$ factors incurred by amplifying the failure probability to $\delta$.
\end{proof}
Finally, we briefly discuss variants of this algorithm.
\begin{itemize}
  \item To get the promised bound in the theorem statement, we can repeatedly estimate $A_iS^T$ (creating $\est_1$, $\est_2$, etc.) with exponentially decaying $\eps$, eventually reducing the error of the first step to $O(\eps\|A_iS^T\|)$.
  This procedure decreases the total variation error to $O(\eps)$ and increases the runtime by $\tilde{O}(\|A_i\|^2/\|D_i\|^2)$, as desired.
  Further, we can ignore $\delta$ by choosing $\delta = \eps$ and outputting an arbitrary $s \in [n]$ upon failure.
  This only changes the output distribution by $\eps$ in total variation distance and increase runtime by $\polylog\frac1\eps$.

  \item While the input is a row $i \in [m]$ (and thus supports query and sampling access), it need not be.
  More generally, given query and sample access to orthonormal vectors $V \in \BB{R}^{n\times k}$, and query access to $x \in \BB{R}^n$, one can approximately sample from a distribution $O(\eps\|x\|/\|VV^Tx\|)$-close to $\CC{D}_{VV^Tx}$, the projection of $x$ onto the span of $V$, in $O(\frac{k^2}{\eps^2}\log\frac{k}{\delta})$ time.

  \item While the SVD dominates the time complexity of Algorithm~\ref{alg:full}, the same description output by \SC{ModFKV} can be used for multiple recommendations, amortizing the cost down to the query complexity (since the rest of the algorithm is linear in the number of queries).
\end{itemize}

\section{Application to Recommendations} \label{sec:model}

We now go through the relevant assumptions necessary to apply Theorem~\ref{main-result-a} to the recommendation systems context.
As mentioned above, these are the same assumptions as those in \cite{kerenidis2016quantum}: an exposition of these assumptions is also given there.
Then, we prove Theorem~\ref{main-result-b}, which shows that Algorithm~\ref{alg:full} gives the same guarantees on recommendations as the quantum algorithm.

\subsection{Preference Matrix} \label{subsec:preference}
Recall that given $m$ users and $n$ products, the preference matrix $T \in \BB{R}^{m\times n}$ contains the complete information on user-product preferences.
For ease of exposition, we will assume the input data is binary:
\begin{definition}
  If user $i$ likes product $j$, then $T_{ij} = 1$.
  If not, $T_{ij} = 0$.
\end{definition}
We can form such a preference matrix from generic data about recommendations, simply by condensing information down to the binary question of whether a product is a good recommendation or not.\footnote{This algorithm makes no distinction between binary matrices and matrices with values in the interval $[0,1]$, and the corresponding analysis is straightforward upon defining a metric for success when data is nonbinary.}

We are typically given only a small subsample of entries of $T$ (which we learn when a user purchases or otherwise interacts with a product).
Then, finding recommendations for user $i$ is equivalent to finding large entries of the $i$th row of $T$ given such a subsample.

Obviously, without any restrictions on what $T$ looks like, this problem is ill-posed.
We make this problem tractable by assuming that $T$ is close to a matrix of small rank $k$.

{\bf $\B{T}$ is close to a low-rank matrix.} That is, $\|T-T_k\|_F \leq \rho\|T\|_F$ for some $k$ and $\rho \ll 1$.
$k$ should be thought of as constant (at worst, $\polylog(m,n)$).
This standard assumption comes from the intuition that users decide their preference for products based on a small number of factors (e.g.\ price, quality, and popularity) \cite{drineas2002competitive,azar2001spectral,koren2009matrix}.

The low-rank assumption gives $T$ robust structure; that is, only given a small number of entries, $T$ can be reconstructed fairly well.

{\bf Many users have approximately the same number of preferences.}
The low-rank assumption is enough to get some bound on quality of recommendations (see Lemma~3.2 in \cite{kerenidis2016quantum}).
However, this bound considers ``matrix-wide'' recommendations.
We would like to give a bound on the probability that an output is a good recommendation {\em for a particular user}.

It is not enough to assume that $\|T-T_k\|_F \leq \rho\|T\|_F$.
In a worst-case scenario, a few users make up the vast majority of the recommendations (say, a few users like every product, and the rest of the users are only happy with four products).
Then, even if we reconstruct $T_k$ exactly, the resulting error, $\rho\|T\|_F$, can exceed the mass of recommendations in the non-heavy users, drowning out any possible information about the vast majority of users that could be gained from the low-rank structure.

In addition to being pathological for user-specific bounds, this scenario is orthogonal to our primary concerns: we aren't interested in providing recommendations to users who desire very few products or who desire nearly all products, since doing so is intractable and trivial, respectively.
To avoid considering such a pathological case, we restrict our attention to the ``typical user'':
\begin{definition}
  For $T \in \BB{R}^{m\times n}$, call $S \subset [m]$ a subset of users {\em $(\gamma, \zeta)$-typical} (where $\gamma > 0$ and $\zeta \in [0,1)$) if $|S| \geq (1-\zeta)m$ and, for all $i \in S$,

    \[ \frac{1}{1+\gamma} \frac{\|T\|_F^2}{m} \leq \|T_i\|^2 \leq (1+\gamma)\frac{\|T\|_F^2}{m}.\]
\end{definition}
$\gamma$ and $\zeta$ can be chosen as desired to broaden or restrict our idea of typical.
We can enforce good values of $\gamma$ and $\zeta$ simply by requiring that users have the same number of good recommendations; this can be done by defining a good recommendation to be the top 100 products for a user, regardless of utility to the user.

Given this definition, we can give a guarantee on recommendations for typical users that come from an approximate reconstruction of $T$.
\begin{theorem} \label{thm:approx-to-sample}
  For $T \in \BB{R}^{m\times n}$, $S$ a $(\gamma,\zeta)$-typical set of users, and a matrix $\tilde{T}$ satisfying $\|T - \tilde{T}\|_F \leq \eps\|T\|_F$,
    \[ \underset{i\sim_u S}{\operatorname{E}}\left[\|\CC{D}_{T_i} - \CC{D}_{\tilde{T}_i}\|_{TV}\right] \leq \frac{2\eps\sqrt{1+\gamma}}{1-\zeta}. \]
  Further, for a chosen parameter $\psi \in (0,1-\zeta)$ there exists some $S' \subset S$ of size at least $(1-\psi-\zeta)m$ such that, for $i \in S'$,
    \[ \|\CC{D}_{T_i} - \CC{D}_{\tilde{T}_i}\|_{TV} \leq 2\eps\sqrt{\frac{1+\gamma}{\psi}}. \]
\end{theorem}
The first bound is an average-case bound on typical users and the second is a strengthening of the resulting Markov bound.
Both bound total variation distance from $\CC{D}_{T_i}$, which we deem a good goal distribution to sample from for recommendations\footnote{When $T$ is not binary, this means that if product $X$ is $\lambda$ times more preferable than product $Y$, then it will be chosen as a recommendation $\lambda^2$ times more often. By changing how we map preference data to actual values in $T$, this ratio can be increased. That way, we have a better chance of selecting the best recommendations, which approaches like matrix completion can achieve. However, these transformations must also preserve that $T$ is close-to-low-rank.}.
We defer the proof of this theorem to the appendix.

When we don't aim for a particular distribution and only want to bound the probability of giving a bad recommendation, we can prove a stronger average-case bound on the failure probability.
\begin{theorem}[Theorem 3.3 of \cite{kerenidis2016quantum}] \label{thm:approx-to-sample2}
  For $T \in \BB{R}^{m\times n}$ a {\em binary} preference matrix, $S$ a $(\gamma,\zeta)$-typical set of users, and a matrix $\tilde{T}$ satisfying $\|T - \tilde{T}\|_F \leq \eps\|T\|_F$, for a chosen parameter $\psi \in (0,1-\zeta)$ there exists some $S' \subset S$ of size at least $(1-\psi-\zeta)m$ such that
    \[ \Pr_{\substack{i \sim_u S' \\ j \sim \tilde{T}_i}} [(i,j) \text{ is bad}] \leq \frac{\eps^2(1+\eps)^2}{(1-\eps)^2\left(1/\sqrt{1+\gamma}-\eps/\sqrt{\psi}\right)^2(1-\psi-\zeta)}. \]
\end{theorem}
For intuition, if $\eps$ is sufficiently small compared to the other parameters, this bound becomes $O(\eps^2(1+\gamma)/(1-\psi-\zeta))$.
The total variation bound from Theorem~\ref{thm:approx-to-sample} is not strong enough to prove this: the failure probability we would get is $2\eps\sqrt{1+\gamma}/(1-\psi-\zeta)$.
Accounting for $T$ being binary gives the extra $\eps$ factor.

{\bf We know $\B{k}$.} More accurately, a rough upper bound for $k$ will suffice.
Such an upper bound can be guessed and tuned from data.

In summary, we have reduced the problem of ``find a good recommendation for a user'' to ``given some entries from a close-to-low-rank matrix $T$, sample from $\tilde{T}_i$ for some $\tilde{T}$ satisfying $\|T-\tilde{T}\|_F \leq \eps\|T\|_F$ for small $\eps$.''

\subsection{Matrix Sampling}
We have stated our assumptions on the full preference matrix $T$, but we also need assumptions on the information we are given about $T$.
Even though we will assume we have a constant fraction of data about $T$, this does not suffice for good recommendations.
For example, if we are given the product-preference data for only half of our products, we have no hope of giving good recommendations for the other half.

We will use a model for subsampling for matrix reconstruction given by Achlioptas and McSherry \cite{achlioptas2007fast}.
In this model, the entries we are given are chosen uniformly over all entries.
This model has seen use previously in the theoretical recommendation systems literature \cite{drineas2002competitive}.
Specifically, we have the following:
\begin{definition}
  For a matrix $T \in \BB{R}^{m\times n}$, let $\hat{T}$ be a random matrix i.i.d.\ on its entries, where
  \begin{equation}\label{eqn:rmatrix}
    \hat{T}_{ij} = \begin{cases} \frac{T_{ij}}{p} & \text{with probability } p \\ 0 & \text{with probability } 1-p\end{cases}. \tag{$\clubsuit$}
  \end{equation}
\end{definition}
Notice that $E[\hat{T}] = T$.
When the entries of $T$ are bounded, $\hat{T}$ is $T$ perturbed by a random matrix $E$ whose entries are independent and bounded random variables.
Standard concentration inequalities imply that such random matrices don't have large singular values (the largest singular value is, say, $O(\sqrt{n/p})$).
Thus, for some vector $v$, if $\|Tv\|/\|v\|$ is large (say, $O(\sqrt{mn/k})$), then $\|(T+E)v\|/\|v\|$ will still be large, despite $E$ having large Frobenius norm.

The above intuition suggests that when $T$ has large singular values, its low-rank approximation $T_k$ is not perturbed much by $E$, and thus, low-rank approximations of $\hat{T}$ are good reconstructions of $T$.
A series of theorems by Achlioptas and McSherry \cite{achlioptas2007fast} and Kerenidis and Prakash \cite{kerenidis2016quantum} formalizes this intuition.
For brevity, we only describe a simplified form of the last theorem in this series, which is the version they (and we) use for analysis.
It states that, under appropriate circumstances, it's enough to compute $\hat{T}_{\sigma,\eta}$ for appropriate $\sigma$ and $\eta$.

\begin{theorem}[4.3 of \cite{kerenidis2016quantum}] \label{thm:lowrank-to-approx-tech}
  Let $T \in \BB{R}^{m\times n}$ and let $\hat{T}$ be the random matrix defined in (\ref{eqn:rmatrix}), with $p \geq \frac{3\sqrt{nk}}{2^{9/2}\eps^3\|T\|_F}$ and $\max_{ij} |T_{ij}| = 1$.
  Let $\sigma = \frac56\sqrt{\frac{\eps^2p}{8k}}\|\hat{T}\|_F$, let $\eta = 1/5$, and assume that $\|T\|_F \geq \frac{9}{\sqrt{2}\eps^3}\sqrt{nk}$.
  Then with probability at least $1-\exp(-19(\log n)^4)$,
    \[ \|T-\hat{T}_{\sigma,\eta}\|_F \leq 3\|T-T_k\|_F + 3\eps\|T\|_F. \]
\end{theorem}

With this theorem, we have a formal goal for a recommendation systems algorithm.
We are given some subsample $A = \hat{T}$ of the preference matrix, along with knowledge of the size of the subsample $p$, the rank of the preference matrix $k$, and an error parameter $\eps$.
Given that the input satisfies the premises for Theorem~\ref{thm:lowrank-to-approx-tech}, for some user $i$, we can provide a recommendation by sampling from $(A_{\sigma,\eta})_i$ with $\sigma,\eta$ specified as described.
Using the result of this theorem, $A_{\sigma,\eta}$ is close to $T$, and thus we can use the results of Section~\ref{subsec:preference} to conclude that such a sample is likely to be a good recommendation for typical users.

Now, all we need is an algorithm that can sample from $(A_{\sigma,\eta})_i$.
Theorem~\ref{main-result-a} shows that Algorithm~\ref{alg:full} is exactly what we need!

\subsection{Proof of Theorem 2} \label{subsec:mrb}

\begin{mr} \label{main-result-b}
  \mrb
\end{mr}

Kerenidis and Prakash's version of this analysis treats $\gamma$ and $\zeta$ with slightly more care, but does eventually assert that these are constants.
Notice that we must assume $p$ is constant.

\begin{proof}
We just run Algorithm~\ref{alg:full} with parameters as described in Proposition~\ref{thm:lowrank-to-approx-tech}: $\sigma = \frac56\sqrt{\frac{\eps^2p}{8k}}\|A\|_F$, $\eps$, $\eta = 1/5$.
Provided $\eps \lesssim \sqrt{p/k}$, the result from Theorem~\ref{main-result-a} holds.
We can perform the algorithm because $A$ is in the data structure given by Proposition~\ref{prop:ds-matrix} (inflating the runtime by a factor of $\log(mn)$).

{\bf Correctness:}
Using Theorem~\ref{thm:lowrank-to-approx-tech} and Theorem~\ref{main-result-a},
\begin{align*}
  \|T-D\|_F &\leq \|T-A_{\sigma,\eta}\|_F + \|A_{\sigma,\eta} - D\|_F \\
  &\leq 3\|T-T_k\|_F + 3\eps\|T\|_F + \eps\|A\|_F.
\end{align*}
Applying a Chernoff bound to $\|A\|_F^2$ (a sum of independent random variables), we get that with high probability $1-e^{-\|T\|_F^2p/3}$, $\|A\|_F \leq \sqrt{2/p}\|T\|_F$.
Since $p$ is constant and $\|T-T_k\|_F \leq \rho\|T\|_F$, we get that $\|T-D\|_F = O(\rho + \eps)\|T\|_F$.

Then, we can apply Theorem~\ref{thm:approx-to-sample} to get that, for $S$ a $(\gamma,\zeta)$-typical set of users of $T$, and $\CC{O}_i$ the output distribution for user $i$, there is some $S' \subset S$ of size at least $(1-\zeta - \psi)m$ such that, for all $i \in S'$,
\begin{align*}
  \|\CC{O}_i-\CC{D}_{T_i}\|_{TV} &\leq \|\CC{O}_i-\CC{D}_{D_i}\|_{TV} + \|\CC{D}_{D_i}-\CC{D}_{T_i}\|_{TV}
  \lesssim \eps + (\eps + \rho)\sqrt{\frac{1+\gamma}{\psi}}
  \lesssim \eps + \rho,
\end{align*}
which is the same bound that Kerenidis and Prakash achieve.

We can also get the same bound that they get when applying Theorem~\ref{thm:approx-to-sample2}: although our total variation error is $\eps$, we can still achieve the same desired $O(\eps^2)$ failure probability as in the theorem.
To see this, notice that in this model, Algorithm~\ref{alg:full} samples from a vector $\alpha$ such that $\|\alpha - T_i\| \leq \eps$.
Instead of using Lemma~\ref{lem:2norm-to-TV}, we can observe that because $T_i$ is binary, the probability that an $\ell^2$-norm sample from $\alpha$ is a bad recommendation is not $\eps$, but $O(\eps^2)$.
From there, everything else follows similarly.

In summary, the classical algorithm has two forms of error that the quantum algorithm does not.
However, the error in estimating the low-rank approximation folds into the error between $T$ and $A_{\sigma,\eta}$, and the error in total variation distance folds into the error from sampling from an inexact reconstruction of $T$.
Thus, we can achieve the same bounds.

{\bf Runtime:}
Our algorithm runs in time and query complexity
    \[ O(\poly(k,1/\eps, \|A_i\|/\|D_i\|)\polylog(mn,1/\delta)), \]
which is the same runtime as Kerenidis and Prakash's algorithm up to polynomial slowdown.

To achieve the stated runtime, it suffices to show that $\|A_i\|/\|D_i\|$ is constant for a constant fraction of users in $S$.
We sketch the proof here; the details are in Kerenidis and Prakash's proof \cite{kerenidis2016quantum}.
We know that $\|T - D\|_F \leq O(\rho + \eps)\|T\|_F$.
Through counting arguments we can show that, for a $(1-\psi')$-fraction of typical users $S'' \subset S$,
    \[ \underset{i \sim_u S''}{\operatorname{E}}\left[\frac{\|A_i\|^2}{\|(A_{\sigma,\eta})_i\|^2}\right] \lesssim \frac{(1+\rho+\eps)^2}{(1-\psi-\zeta)\big(\frac{1}{\sqrt{1+\gamma}}-\frac{\rho+\eps}{\sqrt{\psi}}\big)^2}. \]
For $\rho$ sufficiently small, this is a constant, and so by Markov's inequality a constant fraction $S'''$ of $S''$ has $\|A_i\|/\|D_i\|$ constant.
We choose $\bar{S}$ to be the intersection of $S'''$ with $S'$.
\end{proof}

\section*{Acknowledgments}
Thanks to Scott Aaronson for introducing me to this problem, advising me during the research process, and rooting for me every step of the way.
His mentorship and help were integral to this work as well as to my growth as a CS researcher, and for this I am deeply grateful.
Thanks also to Daniel Liang for providing frequent, useful discussions and for reading through a draft of this document.
Quite a few of the insights in this document were generated during discussions with him.

Thanks to Patrick Rall for the continuing help throughout the research process and the particularly incisive editing feedback.
Thanks to everybody who attended my informal presentations and gave me helpful insight at Simons, including András Gilyén, Iordanis Kerenidis, Anupam Prakash, Mario Szegedy, and Ronald de Wolf.
Thanks to Fred Zhang for pointing out a paper with relevant ideas for future work.
Thanks to Sujit Rao and anybody else that I had enlightening conversations with over the course of the project.
Thanks to Prabhat Nagarajan for the continuing support.

\bibliographystyle{alpha}
{\footnotesize \bibliography{recQuery}}

\newcommand{\etalchar}[1]{$^{#1}$}
\begin{thebibliography}{APSPT05}

\bibitem[Aar15]{aaronson2015read}
Scott Aaronson.
\newblock Read the fine print.
\newblock {\em Nature Physics}, 11(4):291, 2015.

\bibitem[AFK{\etalchar{+}}01]{azar2001spectral}
Yossi Azar, Amos Fiat, Anna Karlin, Frank McSherry, and Jared Saia.
\newblock Spectral analysis of data.
\newblock In {\em Symposium on Theory of Computing}. ACM, 2001.

\bibitem[AM07]{achlioptas2007fast}
Dimitris Achlioptas and Frank McSherry.
\newblock Fast computation of low-rank matrix approximations.
\newblock {\em Journal of the ACM (JACM)}, 54(2):9, 2007.

\bibitem[APSPT05]{Awerbuch:2005:IRS:1070432.1070599}
Baruch Awerbuch, Boaz Patt-Shamir, David Peleg, and Mark Tuttle.
\newblock Improved recommendation systems.
\newblock In {\em Symposium on Discrete Algorithms}, 2005.

\bibitem[BBBV97]{bennett1997strengths}
Charles~H Bennett, Ethan Bernstein, Gilles Brassard, and Umesh Vazirani.
\newblock Strengths and weaknesses of quantum computing.
\newblock {\em SIAM Journal on Computing}, 26(5):1510--1523, 1997.

\bibitem[BJ99]{Bshouty:1999:LDO:305673.305756}
Nader~H. Bshouty and Jeffrey~C. Jackson.
\newblock Learning {DNF} over the uniform distribution using a quantum example
  oracle.
\newblock {\em SIAM J. Comput.}, 28(3):1136--1153, 1999.

\bibitem[BK07]{bell2007lessons}
Robert~M Bell and Yehuda Koren.
\newblock Lessons from the {Netflix} prize challenge.
\newblock {\em ACM SIGKDD Explorations Newsletter}, 9(2):75--79, 2007.

\bibitem[DKM06]{doi:10.1137/S0097539704442684}
P.~Drineas, R.~Kannan, and M.~Mahoney.
\newblock Fast {Monte} {Carlo} algorithms for matrices {I}: Approximating
  matrix multiplication.
\newblock {\em SIAM Journal on Computing}, 36(1):132--157, 2006.

\bibitem[DKR02]{drineas2002competitive}
Petros Drineas, Iordanis Kerenidis, and Prabhakar Raghavan.
\newblock Competitive recommendation systems.
\newblock In {\em Symposium on Theory of Computing}. ACM, 2002.

\bibitem[DMM08]{doi:10.1137/07070471X}
P.~Drineas, M.~Mahoney, and S.~Muthukrishnan.
\newblock Relative-error {CUR} matrix decompositions.
\newblock {\em SIAM Journal on Matrix Analysis and Applications},
  30(2):844--881, 2008.

\bibitem[DV06]{deshpande2006adaptive}
Amit Deshpande and Santosh Vempala.
\newblock Adaptive sampling and fast low-rank matrix approximation.
\newblock In {\em Approximation, Randomization, and Combinatorial Optimization.
  Algorithms and Techniques}, pages 292--303. Springer, 2006.

\bibitem[FKV04]{frieze2004fast}
Alan Frieze, Ravi Kannan, and Santosh Vempala.
\newblock Fast {Monte-Carlo} algorithms for finding low-rank approximations.
\newblock {\em Journal of the ACM (JACM)}, 51(6):1025--1041, 2004.

\bibitem[GSLW18]{gilyen2018quantum}
Andr{\'a}s Gily{\'e}n, Yuan Su, Guang~Hao Low, and Nathan Wiebe.
\newblock Quantum singular value transformation and beyond: exponential
  improvements for quantum matrix arithmetics.
\newblock {\em arXiv}, 2018.

\bibitem[HHL09]{harrow2009quantum}
Aram~W Harrow, Avinatan Hassidim, and Seth Lloyd.
\newblock Quantum algorithm for linear systems of equations.
\newblock {\em Physical review letters}, 103(15):150502, 2009.

\bibitem[HKS11]{hazan2011beating}
Elad Hazan, Tomer Koren, and Nati Srebro.
\newblock Beating {SGD}: Learning {SVMs} in sublinear time.
\newblock In {\em Neural Information Processing Systems}, 2011.

\bibitem[KBV09]{koren2009matrix}
Yehuda Koren, Robert Bell, and Chris Volinsky.
\newblock Matrix factorization techniques for recommender systems.
\newblock {\em Computer}, 42(8), 2009.

\bibitem[KP17a]{kerenidis2017quantum}
Iordanis Kerenidis and Anupam Prakash.
\newblock Quantum gradient descent for linear systems and least squares.
\newblock {\em arXiv}, 2017.

\bibitem[KP17b]{kerenidis2016quantum}
Iordanis Kerenidis and Anupam Prakash.
\newblock Quantum recommendation systems.
\newblock In {\em Innovations in Theoretical Computer Science}, 2017.

\bibitem[KRRT01]{KUMAR200142}
Ravi Kumar, Prabhakar Raghavan, Sridhar Rajagopalan, and Andrew Tomkins.
\newblock Recommendation systems: a probabilistic analysis.
\newblock {\em Journal of Computer and System Sciences}, 63(1):42 -- 61, 2001.

\bibitem[KS08]{kleinberg2008using}
Jon Kleinberg and Mark Sandler.
\newblock Using mixture models for collaborative filtering.
\newblock {\em Journal of Computer and System Sciences}, 74(1):49--69, 2008.

\bibitem[KV17]{kannan_vempala_2017}
Ravindran Kannan and Santosh Vempala.
\newblock Randomized algorithms in numerical linear algebra.
\newblock {\em Acta Numerica}, 26:95–135, 2017.

\bibitem[LMR13]{lloyd2013quantum}
Seth Lloyd, Masoud Mohseni, and Patrick Rebentrost.
\newblock Quantum algorithms for supervised and unsupervised machine learning.
\newblock {\em arXiv}, 2013.

\bibitem[LMR14]{lloyd2014quantum}
Seth Lloyd, Masoud Mohseni, and Patrick Rebentrost.
\newblock Quantum principal component analysis.
\newblock {\em Nature Physics}, 10(9):631, 2014.

\bibitem[Pre18]{preskill2018quantum}
John Preskill.
\newblock Quantum {Computing} in the {NISQ} era and beyond.
\newblock {\em {Quantum}}, 2:79, 2018.

\bibitem[Rec11]{recht2011simpler}
Benjamin Recht.
\newblock A simpler approach to matrix completion.
\newblock {\em Journal of Machine Learning Research}, 12:3413--3430, 2011.

\bibitem[SWZ16]{song2016sublinear}
Zhao Song, David~P. Woodruff, and Huan Zhang.
\newblock Sublinear time orthogonal tensor decomposition.
\newblock In {\em Neural Information Processing Systems}, 2016.

\end{thebibliography}
\addcontentsline{toc}{section}{References}

\appendix
\section{Deferred Proofs}

\begin{proof}[Proof of Lemma~\ref{lem:fkvbound}]
    We can describe \SC{ModFKV} as FKV run on $K$ with the filter threshold $\gamma$ raised from $\Theta(\bar{\eps}/K)$ to $1/K$.
    The original work aims to output a low-rank approximation similar in quality to $A_K$, so it needs to know about singular values as low as $\bar{\eps}/K$.
    In our case, we don't need as strong of a bound, and can get away with ignoring these singular vectors.
    To prove our bounds, we just discuss where our proof differs from the original work's proof (Theorem~1 of \cite{frieze2004fast}).
    First, they show that
      \[ \Delta(W^T; u^{(t)},t \in [K]) \geq \|A_K\|_F^2 - \frac{\bar{\eps}}{2}\|A\|_F^2. \]
    The proof of this holds when replacing $K$ with any $K' \leq K$.
    We choose to replace $K$ with the number of singular vectors taken by \SC{ModFKV}, $k$.
    Then we have that
      \[ \Delta(W^T; u^{(t)}, t \in T) = \Delta(W^T; u^{(t)},t \in [k]) \geq \|A_k\|_F^2 - \frac{\bar{\eps}}{2}\|A\|_F^2. \]
    We can complete the proof now, using that $[k] = T$ because our filter accepts the top $k$ singular vectors (though not the top $K$).
    Namely, we avoid the loss of $\gamma\|W\|_F^2$ that they incur in this way.
    This gives the bound (\ref{eqn:fkvbound}).

    Further, because we raise $\gamma$, we can correspondingly lower our number of samples.
    Their analysis requires $q$ (which they denote $p$) to be $\Omega(\max\{\frac{k^4}{\bar{\eps}^2}, \frac{k^2}{\bar{\eps}\gamma^2}, \frac{1}{\bar{\eps}^2\gamma^2}\})$ (for Lemma~3, Claim~1, and Claim~2, respectively).
    So, we can pick $q = \Theta(K^4/\bar{\eps}^2)$.

    As for bounding $k$, \SC{ModFKV} can compute the first $k$ singular values to within a cumulative additive error of $\bar{\eps}\|A\|_F$.
    This follows from Lemma~2 of \cite{frieze2004fast} and the Hoffman-Wielandt inequality.
    Thus, \SC{ModFKV} could only conceivably take a singular vector $v$ such that $\|Av\| \geq \sigma - \bar{\eps}\|A\|_F = \sigma(1 - \bar{\eps}\|A\|_F/\sigma)$, and analogously for the upper bound.
\end{proof}

\begin{proof}[Proof of Proposition~\ref{prop:modfkv-apporth}]
  We follow the proof of Claim~2 of \cite{frieze2004fast}.
  For $i \neq j$, we have as follows:
  \begin{gather*}
    \big|\hat{v}_i^T\hat{v}_j\big| = \frac{|u_i^TSS^Tu_j|}{\|W^Tu_i\|\|W^Tu_j\|} \leq
    \frac{|u_i^TSS^Tu_j|}{\sigma^2} \leq \frac{\|S\|_F^2}{\sigma^2\sqrt{q}} = \frac{\bar{\eps}}{K}\\
    \big|1 - \hat{v}_i^T\hat{v}_i\big|
    = \frac{|u_i^TWW^Tu_i| - |u_i^TSS^Tu_i|}{\|W^Tu_i\|\|W^Tu_i\|} \leq
    \frac{\|S\|_F^2}{\sigma^2\sqrt{q}} = \frac{\bar{\eps}}{K}
  \end{gather*}
  Here, we use that $\|WW^T - SS^T\| \leq \|S\|_F^2/\sqrt{q}$ (Lemma~2 \cite{frieze2004fast}) and $\{Wu_i\}$ are orthogonal.

  This means that $\hat{V}^T\hat{V}$ is $O(\bar{\eps}/K)$-close entry-wise to the identity.
  Looking at $\hat{V}$'s singular value decomposition into $A\Sigma B^T$ (treating $\Sigma$ as square), the entrywise bound implies that $\|\Sigma^2 - I\|_F \lesssim \bar{\eps}$, which in turn implies that $\|\Sigma - I\|_F \lesssim \bar{\eps}$.
  $\Lambda := AB^T$ is close to $\hat{V}$, orthonormal, and in the same subspace as desired.
\end{proof}

\begin{proof}[Proof of Theorem~\ref{thm:bound-transfer}]
  We will prove a slightly stronger statement: it suffices to choose $A_{\sigma,\eta}$ such that $P_{\sigma,\eta}$ is an orthogonal projector\footnote{In fact, we could have used this restricted version as our definition of $A_{\sigma,\eta}$.} (denoted $\Pi_{\sigma,\eta}$).
  We use the notation $\Pi_E := \Pi_{\sigma,\eta} - \Pi_{ \sigma(1+\eta)}$ to refer to the error of $\Pi_{\sigma,\eta}$, which can be any orthogonal projector on the span of the singular vectors with values in $[\sigma(1-\eta), \sigma(1+\eta))$.
  We denote $\sigma_k$ by $\sigma$ and $\min m,n$ by $N$.
  \begin{align*}
    \|A\Pi - A_{\sigma,\eta}\|_F^2 &= \|U\Sigma V^T(\Pi - \Pi_{\sigma,\eta})\|_F^2 \\
    &= \|\Sigma V^T(\Pi - \Pi_{\sigma,\eta})\|_F^2 \\
    &= \sum_{i=1}^N\sigma_i^2\|v_i^T\Pi - v_i^T\Pi_{\sigma,\eta}\|^2
  \end{align*}
  That is, $A\Pi$ and $A_{\sigma,\eta}$ are close when their corresponding projectors behave in the same way.
  Let $a_i = v_i^T\Pi$, and $b_i = v_i^T\Pi_{\sigma,\eta}$.
  Note that
    \[ b_i = \begin{cases}
      v_i^T & \sigma_i \geq (1+\eta)\sigma \\
      v_i^T\Pi_E & (1+\eta)\sigma > \sigma_i \geq (1-\eta)\sigma \\
      0 & (1-\eta)\sigma > \sigma_i
    \end{cases}. \]
  Using the first and third case, and the fact that orthogonal projectors $\Pi$ satisfy $\|v - \Pi v\|^2 = \|v\|^2 - \|\Pi v\|^2$, the formula becomes
  \begin{multline} \label{eqn:lramain}
    \|A\Pi - A_{\sigma,\eta}\|_F^2 =
    \sum_{1}^{\ell(\sigma(1+\eta))} \sigma_i^2(1-\|a_i\|^2)
    + \sum_{\ell(\sigma(1+\eta))+1}^{\ell(\sigma(1-\eta))} \sigma_i^2\|a_i - b_i\|^2
    + \sum_{\ell(\sigma(1-\eta))+1}^N\sigma_i^2(\|a_i\|^2). \tag{$\spadesuit$}
  \end{multline}
  Now, we consider the assumption equation.
  We reformulate the assumption into the following system of equations:
  \begin{gather*}
    \sum_{i=1}^k \sigma_i^2 \leq \sum_{i=1}^N \sigma_i^2\|a_i\|^2 + \eps\sigma^2 \qquad
    \sigma_i^2 \text{ are nonincreasing} \\
    \|a_i\|^2 \in [0,1] \qquad
    \sum \|a_i\|^2 = k
  \end{gather*}
  The first line comes from the equation.
  The second line follows from $\Pi$ being an orthogonal projector on a $k$-dimensional subspace.

  It turns out that this system of equations is enough to show that the $\|a_i\|^2$ behave the way we want them to.
  We defer the details to Lemma~\ref{lem:system-solution}; the results are as follows.
  \begin{align*}
    \sum_1^{\ell(\sigma_k(1+\eta))}\sigma_i^2(1-\|a_i\|^2) &\leq \eps\Big(1+\frac{1}{\eta}\Big)\sigma_k^2 &
    \sum_{\ell(\sigma_k(1-\eta))+1}^N \sigma_i^2\|a_i\|^2 &\leq \eps\Big(\frac{1}{\eta}-1\Big)\sigma_k^2 \\
    \sum_1^{\ell(\sigma_k1+\eta))}(1-\|a_i\|^2) &\leq \frac{\eps}{\eta} &
    \sum_{\ell(\sigma_k(1-\eta))+1}^N \|a_i\|^2 &\leq \frac{\eps}{\eta}
  \end{align*}
  Now, applying the top inequalities to (\ref{eqn:lramain}):
    \[ \|A\Pi - A_{\sigma,\eta}\|_F^2 \leq \frac{2\eps\sigma^2}{\eta} + \sum_{\ell(\sigma(1+\eta))+1}^{\ell(\sigma(1-\eta))} \sigma_i^2\|a_i - b_i\|^2. \]
  We just need to bound the second term of (\ref{eqn:lramain}).
  Notice the following:
    \[ \sum_{\ell(\sigma(1+\eta))+1}^{\ell(\sigma(1-\eta))} \sigma_i^2\|a_i - b_i\|^2 \leq \sigma^2(1+\eta)^2\|U^T(\Pi-\Pi_E)\|_F^2, \]
  where $U$ is the set of vectors $v_{\ell(\sigma(1+\eta))+1}$ through $v_{\ell(\sigma(1-\eta))}$.

  Notice that $\Pi_E$ is the error component of the projection, and this error can be any projection onto a subspace spanned by $U$.
  Thus, to bound the above we just need to pick an orthogonal projector $\Pi_E$ making the norm as small as possible.
  If $UU^T\Pi$ were an orthogonal projection, this would be easy:
    \[ \|U^T(\Pi - UU^T\Pi)\|_F^2 = 0. \]
  However, this is likely not the case.
  $UU^T\Pi$ is {\em close} to an orthogonal projector, though, through the following reasoning:

  For ease of notation let $P_1$ be the orthogonal projector onto the first $\ell(\sigma(1+\eta))$ singular vectors, $P_2 = UU^T$, and $P_3$ be the orthogonal projector onto the the rest of the singular vectors.
  We are concerned with $P_2\Pi$.

  Notice that $P_1 + P_2 + P_3 = I$.
  Further, $\|(I-\Pi)P_1\|_F^2 \leq \eps/\eta$ and $\|\Pi P_3\|_F^2 \leq \eps/\eta$ from Lemma~\ref{lem:system-solution}.
  Then
  \begin{align*}
    P_2\Pi &= (I-P_1-P_3)\Pi = \Pi - P_1 + P_1(I-\Pi) - P_3\Pi \\
    \|P_2\Pi - (\Pi - P_1)\|_F &= \|P_1(I-\Pi) - P_3\Pi\|_F \leq 2\sqrt{\eps/\eta}
  \end{align*}
  So now it is sufficient to show that $\Pi - P_1$ is close to a projector matrix.
  This follows from Lemma~\ref{lem:hermitian}, since it satisfies the premise:
  \begin{align*}
    (\Pi-P_1)^2 - (\Pi-P_1) &= \Pi-\Pi P_1-P_1\Pi+P_1-\Pi+P_1 \\
    &= (I-\Pi)P_1 + P_1(I-\Pi) \\
    \|(\Pi-P_1)^2 - (\Pi-P_1)\|_F &\leq 2\sqrt{\eps/\eta}
  \end{align*}
  Thus, $UU^T\Pi$ is $(2\sqrt{\eps/\eta} + (2\sqrt{\eps/\eta} + 16\eps/\eta))$-close to an orthogonal projector in Frobenius norm.

  We can choose $\Pi_E$ to be $M$, and plug this into (\ref{eqn:lramain}).
  We use the assumptions that $\eps/\eta < 1$ and $\eta < 1$ to bound.
  \begin{align*}
    \sum_{\ell(\sigma(1+\eta))+1}^{\ell(\sigma(1-\eta))} \sigma_i^2\|a_i - b_i\|^2 &\leq \sigma^2(1+\eta)^2\|U^T(\Pi-M)\|_F^2 \\
    &\leq \sigma^2(1+\eta)^2\|U^T(\Pi-(UU^T\Pi+E))\|_F^2 \\
    &\leq \sigma^2(1+\eta)^2\|U^TE\|_F^2 \\
    &\lesssim \sigma^2(1+\eta)^2\eps/\eta \\
    \|A\Pi - A_{\sigma,\eta}\|_F^2 &\lesssim \frac{2\eps\sigma^2}{\eta} + \sigma^2(1+\eta)^2\frac{\eps}{\eta}
    \lesssim \eps\sigma^2/\eta
  \end{align*}
  This concludes the proof.
  (The constant factor is 1602.)
\end{proof}

\begin{lemma} \label{lem:hermitian}
  If a Hermitian $A$ satisfies $\|A^2-A\|_F \leq \eps$, then $\|A - P\|_F \leq \eps + 4\eps^2$ for some orthogonal projector $P$.
\end{lemma}
\begin{proof}
  Use the fact that Hermitian matrices are normal, so $A = U\Gamma U^T$ for unitary $U$ and diagonal matrix $\Gamma$, and
    \[ A^2 - A = U(\Gamma^2 - \Gamma)U^T \implies \|\Gamma^2 - \Gamma\|_F \leq \eps .\]
  From here, consider the entries $\gamma_i$ of $\Gamma$, satisfying $\gamma_i^2 - \gamma_i = c_i$ and $\sum c_i^2 = \eps^2$.
  Thus, $\gamma_i = (1\pm \sqrt{1+4c_i})/2$ which is at most $c_i + 4c_i^2$ off from $0.5 \pm 0.5$ (aka $\{0,1\}$), using that 
    \[ 1-x/2-x^2/2 \leq \sqrt{1-x} \leq 1-x/2. \]
  Finally, this means that $\Gamma$ is off from having only 0's and 1's on the diagonal by $\sqrt{\sum (c_i + 4c_i^2)^2} \leq \eps + 4\eps^2$ in Frobenius norm.
  If $\Gamma$ had only 0's and 1's on the diagonal, the resulting $U\Gamma U^T$ would be an orthogonal projector.
\end{proof}

\begin{lemma} \label{lem:system-solution}
  The system of equations:
  \begin{gather*}
    \sum_{i=1}^k \sigma_i^2 \leq \sum_{i=1}^N \sigma_i^2\|a_i\|^2 + \eps\sigma_k^2 \qquad
    \sigma_i^2 \text{ are nonincreasing} \\
    \|a_i\|^2 \in [0,1] \qquad
    \sum \|a_i\|^2 = k
  \end{gather*}
  imply the following, for $0 < \eta \leq 1$:
  \begin{align*}
    \sum_1^{\ell(\sigma_k(1+\eta))}\sigma_i^2(1-\|a_i\|^2) &\leq \eps\Big(1+\frac{1}{\eta}\Big)\sigma_k^2 &
    \sum_{\ell(\sigma_k(1-\eta))+1}^N \sigma_i^2\|a_i\|^2 &\leq \eps\Big(\frac{1}{\eta}-1\Big)\sigma_k^2 \\
    \sum_1^{\ell(\sigma_k1+\eta))}(1-\|a_i\|^2) &\leq \frac{\eps}{\eta} &
    \sum_{\ell(\sigma_k(1-\eta))+1}^N \|a_i\|^2 &\leq \frac{\eps}{\eta}
  \end{align*}
\end{lemma}
\begin{proof}
  We are just proving straightforward bounds on a linear system.
  We will continue to denote $\sigma_k$ by $\sigma$.
  Thus, $k = \ell(\sigma)$.

  The slack in the inequality is always maximized when the weight of the $\|a_i\|^2$ is concentrated on the large-value (small-index) entries.
  For example, the choice of $\|a_i\|^2$ maximizing slack in the given system of equations is the vector $ \{\|a_i\|\}_{i \in [N]} = \B{1}_{\leq k}$.
  Here, $\B{1}_{\leq x}$ denotes the vector where
    \[ \left(\B{1}_{\leq x}\right)_{i} := \begin{cases} 1 & i \leq x \\ 0 & \text{otherwise} \end{cases}. \]

  For brevity, we only give the details for the first bound; the others follow similarly.
  Consider adding the constraint $C = \sum_1^{\ell(\sigma(1+\eta))}\sigma_i^2(1-\|a_i\|^2)$ to the system of equations.
  We want to determine for which values of $C$ the modified system is still feasible; we can do this by trying the values that maximize slack.

  This occurs when weight is on the smallest possible indices: when $\|a_{\ell(\sigma(1+\eta))}\|^2 = 1-C/\sigma_{\ell(\sigma(1+\eta))}^2$, $\|a_{\ell(\sigma)+1}\|^2 = C/\sigma_{\ell(\sigma(1+\eta))}^2$, and all other $\|a_i\|^2$ are $\B{1}_{\geq k}$.
  Notice that $\|a_{\ell(\sigma(1+\eta))}\|^2$ could be negative and $\|a_{\ell(\sigma)+1}\|$ could be larger than one, breaking constraints.
  However, if there is no feasible solution even when relaxing those two constraints, there is certainly no solution to the non-relaxed system.
  Thus, we check feasibility (by construction the second equation is satisfied):
  \begin{align*}
    \sum_{i=1}^k \sigma_i^2 &\leq \sum_{i=1}^k \sigma_i^2 - C + C\frac{\sigma_{\ell(\sigma)+1}^2}{\sigma_{\ell(\sigma(1+\eta))}} + \eps\sigma^2 \\
    C\Big(1-\frac{\sigma_{\ell(\sigma)+1}^2}{\sigma_{\ell(\sigma(1+\eta))}}\Big)&\leq \eps\sigma^2 \\
    C\Big(1-\frac{1}{(1+\eta)^2}\Big)&\leq \eps\sigma^2
  \end{align*}
  This gives the bound on $C$.
  Repeating for all four cases, we get the following bounds:
  \begin{align*}
    \sum_1^{\ell(\sigma(1+\eta))}\sigma_i^2(1-\|a_i\|^2) &\leq \frac{\eps(1+\eta)^2\sigma^2}{2\eta+\eta^2} &
    \sum_{\ell(\sigma(1-\eta))+1}^N \sigma_i^2\|a_i\|^2 &\leq \frac{\eps(1-\eta)^2\sigma^2}{2\eta-\eta^2} \\
    \sum_1^{\ell(\sigma(1+\eta))}(1-\|a_i\|^2) &\leq \frac{\eps}{2\eta+\eta^2} &
    \sum_{\ell(\sigma(1-\eta))+1}^N \|a_i\|^2 &\leq \frac{\eps}{2\eta-\eta^2}
  \end{align*}
  We get the bounds in the statement by simplifying the above (using that $\eta \leq 1$).
\end{proof}

\begin{proof}[Proof of Theorem~\ref{thm:approx-to-sample}]
  The following shows the first, average-case bound (note the use of Lemma~\ref{lem:2norm-to-TV} and Cauchy-Schwarz).
  \begin{align*}
    \operatorname{E}_{i \sim_u S}\left[\|\CC{D}_{T_i}- \CC{D}_{\tilde{T}_i}\|_{TV}\right] &= \frac{1}{|S|}\sum_{i \in S} \|\CC{D}_{T_i} - \CC{D}_{\tilde{T}_i}\|_{TV} \\
    &\leq \frac{1}{(1-\zeta)m}\sum_{i \in S} \frac{2\|T_i - \tilde{T}_i\|}{\|T_i\|} \\
    &\leq \frac{2(1+\gamma)}{(1-\zeta)\sqrt{m}\|T\|_F}\sum_{i \in S} \|T_i - \tilde{T}_i\| \\
    &\leq 2\frac{1+\gamma}{1-\zeta}\Big(\frac{\sum_{i \in [m]} \|T_i - \tilde{T}_i\|}{\sqrt{m}\|T\|_F}\Big) \\
    &\leq 2\frac{1+\gamma}{1-\zeta}\Big(\frac{\sqrt{m}\|T-\tilde{T}\|_F}{\sqrt{m}\|T\|_F}\Big) \\
    &\leq \frac{2\eps(1+\gamma)}{(1-\zeta)}
  \end{align*}

  Using that $\|T-\tilde{T}\|_F \leq \eps\|T\|_F$ in combination with a pigeonhole-like argument, we know that at least a $(1-\psi)$-fraction of users $i \in [m]$ satisfy
    \[ \|T_i - \tilde{T}_i\|^2 \leq \frac{\eps^2\|A\|_F^2}{\psi m}. \]
  Thus, there is a $S' \subset S$ of size at least $(1-\psi-\zeta)m$ satisfying the above.
  For such an $i \in S'$, we can argue from Lemma~\ref{lem:2norm-to-TV} and the definition of a $(\gamma,\zeta)$-typical user that
    \[ \|\CC{D}_{T_i} -\CC{D}_{\tilde{T}_i}\|_{TV} \leq \frac{2\|T_i - \tilde{T}_i\|}{\|T_i\|} \leq \frac{2\eps\|T\|_F(1+\gamma)\sqrt{m}}{\sqrt{\psi m}\|T\|_F} = \frac{2\eps(1+\gamma)}{\sqrt{\psi}}. \]
\end{proof}

\section{Variant for an Alternative Model} \label{sec:drineas}

In this section, we describe a variant of our recommendation systems algorithm for the competitive recommendations model, seen in Drineas, Kerenidis, and Raghavan's 2002 paper giving two algorithms for competitive recommendations \cite{drineas2002competitive}.
The idea is to output good recommendations with as little knowledge about the preference matrix $T$ as possible.
Our algorithm is similar to Drineas et al's second algorithm, which has weak assumptions on the form of $T$, but strong assumptions on how we can gain knowledge about it.

We use a similar model, as follows:
\begin{itemize}
  \item We begin with no knowledge of our preference matrix $T$ apart from the promise that $\|T - T_k\|_F \leq \rho\|T\|_F$;
  \item We can request the value of an entry $T_{ij}$ for some cost;
  \item For some constant $0 < c \leq 1$, we can sample from and compute probabilities from a distribution $P$ over $[m]$ satisfying
    \[ P(i) \geq c\frac{\|T_i\|^2}{\|T\|_F^2}. \]
  Further, we can sample from and compute probabilities from distributions $Q_i$ over $[n]$, for $i \in [m]$, satisfying
    \[ Q_i(j) \geq c\frac{T_{ij}^2}{\|T_i\|^2}. \]
\end{itemize}
We discuss the first assumption in Section~\ref{subsec:preference}.
The second assumption is very strong, but we will only need to use it sparingly, for some small set of users and products.
In practice, this assumption could be satisfied through paid user surveys.

The last assumption states that the way that we learn about users naturally, via normal user-site interaction, follows the described distributions.
For example, consider when $T$ is binary (as in Section~\ref{subsec:preference}).
The assumption about $P$ states that we can sample for users proportional to the number of products they like (with possible error via $c$).
Even though we don't know the exact number of products a user likes, it is certainly correlated with the amount of purchases/interactions the user has with the site.
With this data we can form $P$.
The assumption about $Q_i$'s states that, for a user, we can sample uniformly from the products that user likes.
We can certainly assume the ability to sample from the products that a user likes, since such positive interactions are common, intended, and implicit in the user's use of the website.
It is not clear whether uniformity is a reasonable assumption, but this can be made more reasonable by making $T$ non-binary and more descriptive of the utility of products to users.

Under these assumptions, our goal is, given a user $i$, to recommend products to that user {\em that were not previously known to be good} and are likely to be good recommendations.

To do this, we run Algorithm~\ref{alg:full} with $T, k, \eps$ as input, the main change being that we use Frieze, Kannan, and Vempala's algorithm as written in their paper instead of \SC{ModFKV}.
As samples and requests are necessary, we can provide them using the assumptions above.

For the FKV portion of the algorithm, this leads to $O(q^2)$ requests to $q$ users about $q$ products, where $q = O(\max\{\frac{k^4}{c^3\eps^6},\frac{k^2}{c^3\eps^8}\})$.
This gives the description of a $D$ such that
  \[ \|T-D\|_F \leq \sqrt{\|T-T_k\|_F^2 + \eps^2\|T\|_F^2} \leq (\rho+\eps)\|T\|_F. \]
Thus, immediately we can use theorems from Section~\ref{subsec:preference} to show that samples from $D$ will give good recommendations.

From here, the next part of Algorithm~\ref{alg:full} can output the desired approximate sample from $D_i$.
A similar analysis will show that this approximate sample is likely to be a good recommendation, all while requesting and sampling a number of entries independent of $m$ and $n$.
Such requests and samples will only be needed for the $q$ users chosen by FKV for its subsample, along with the input user.
Further, for more recommendations, this process can be iterated with unused information about the $q$ users chosen by FKV.
Alternatively, if we can ask the $q$ users for all of their recommendations, we only need $O(\frac{k^2}{\eps^2}\log \frac{k}{\delta})$ samples from the input user to provide that user with an unlimited number of recommendations (we can store and update the estimate of $A_iS^T$ to use when sampling).

This gives good recommendations, only requiring knowledge of $O(\poly(k,1/\eps))$ entries of $T$, and with time complexity polynomial in the number of known entries.
\end{document}